\documentclass[11pt]{article}

\usepackage[T1]{fontenc}
\usepackage[utf8]{inputenc}
\usepackage{newtxtext,newtxmath}
\usepackage{geometry}
\usepackage{microtype}
\usepackage[hidelinks]{hyperref}
\usepackage{amsmath}
\usepackage[round,authoryear]{natbib}
\usepackage{fvextra}
\DefineVerbatimEnvironment{pseudocode}{Verbatim}{
  breaklines=true,
  breakanywhere=true,
  breaksymbolleft={},
  fontsize=\small,
  numbers=left,
  numbersep=6pt
}

\usepackage{graphicx}
\usepackage{placeins}
\usepackage{makecell}

\newcommand{\colRA}[1]{\makecell[c]{RA-\\nWG\\@#1}}
\newcommand{\colNR}[2]{\makecell[c]{N-Recall\\#1\\@#2}}
\usepackage{float}
\usepackage{adjustbox}
\usepackage{textcomp}

\usepackage[font=small,labelfont=bf]{caption}
\newcommand{\TitleSize}{\fontsize{26}{30}\selectfont}
\newcommand{\AuthorNameSize}{\fontsize{16}{19}\selectfont}
\newcommand{\RuleThickness}{0.6pt}

\hypersetup{
  unicode=true,
  pdfencoding=auto,
  pdftitle  = {Practical RAG Evaluation: A Rarity-Aware Set-Based Metric and Cost-Latency-Quality Trade-offs},
  pdfauthor = {Etienne Dallaire}
}
\begin{document}

% ------------ custom title block with 2 rules ------------
\begin{center}
\rule{\linewidth}{\RuleThickness}\par
\vspace{0.9\baselineskip}

{\TitleSize Practical RAG Evaluation:\\
A Rarity-Aware Set-Based Metric\\
and \mbox{Cost--Latency--Quality} Trade-offs\par}

\vspace{0.9\baselineskip}
\rule{\linewidth}{\RuleThickness}
\end{center}

\vspace{1.1\baselineskip}
\begin{center}
{\AuthorNameSize Etienne Dallaire\par}
\vspace{0.25\baselineskip}
Independent Researcher\\
Paris, France
\end{center}
\vspace{1.0\baselineskip}

% ------------ abstract ------------
\begin{center}\textsc{Abstract}\end{center}
\noindent This paper addresses the guessing game in building production RAG. Classical rank-centric IR metrics (nDCG/MAP/MRR) misfit RAG, where LLMs consume a set of passages rather than a browsed list; position discounts and prevalence-blind aggregation miss what matters: whether the prompt at cutoff K contains the decisive evidence. Second, there is no standardized, reproducible way to build and audit golden sets. Third, leaderboards exist but lack end-to-end, on-corpus benchmarking that reflects production trade-offs. Fourth, how state-of-the-art embedding models handle proper-name identity signals and conversational noise remains opaque. To address these, we contribute: (1) RA-nWG@K, a rarity-aware, per-query-normalized set score, and operational ceilings via the pool-restricted oracle ceiling (PROC) and \%PROC to separate retrieval from ordering headroom within a Cost-Latency-Quality (CLQ) lens; (2) rag-gs (MIT), a lean golden-set pipeline with Plackett-Luce listwise refinement whose iterative updates outperform single-shot LLM ranking; (3) a comprehensive benchmark on a production RAG (scientific-papers corpus) spanning dense retrieval, hybrid dense+BM25, embedding models and dimensions, cross-encoder rerankers, ANN/HNSW, and quantization; and (4) targeted diagnostics that quantify proper-name identity signal and conversational-noise sensitivity via identity-destroying and formatting ablations. Together, these components provide practitioner Pareto guidance and auditable guardrails to support reproducible, budget/SLA-aware decisions.

\clearpage

\section{Introduction \& Motivation}

Production retrieval-augmented generation (RAG) is a set-consumption pipeline: the generator receives a bounded context consisting of the top-\(K\) retrieved passages serialized into a single prompt, and downstream utility depends almost entirely on whether that set contains the decisive evidence under the prompt budget. Rank-centric assumptions inherited from interactive IR---position discounting, smoothness of scores, user browsing---are therefore misaligned with the operational objective, which is to maximize evidence presence and usefulness at fixed \(K\) subject to service-level constraints. Because retrieval decisions jointly determine both quality (which items enter the prompt) and spend/latency (prompt token count, first-token delay, and reranking cost), we adopt a cost--latency--quality (CLQ) lens that treats pre-generation choices---embedder, ANN configuration, candidate depth \(K\), hybridization, and reranking---as controls on a constrained optimization. Generator input cost scales approximately linearly with the total prompt tokens (\(\propto K \times\) tokens per chunk), and first-token latency increases with the same; hence any increase in \(K\) or chunk length simultaneously raises reranking expense and final inference time. Real workloads compound this with label heterogeneity across queries: the available supply of high-utility passages (grade-5/grade-4 under a fixed rubric) varies by orders of magnitude, so unnormalized metrics conflate system behavior with prevalence.

We ground evaluation and operations in three design commitments. First, we evaluate sets, not ranks. Our core metric, RA-nWG@\(K\), is order-free, per-query normalized, and rarity-aware: it aggregates stationary per-passage utilities, scales mid-grades by inverse prevalence under strict caps so grade-5 dominance is preserved, and divides by the query's pool-restricted oracle at \(K\) (PROC) to yield a \([0,1]\) score comparable across heterogeneous label mixes. Second, we pair RA-nWG@\(K\) with coverage diagnostics---N-Recall\(_{4+}\)@\(K\) and N-Recall\(_{5}\)@\(K\)---that report the fraction of available high-utility evidence captured under the same budget; when the deployed generator is order-sensitive or distractible, we supplement with a harm rate (Harm@\(K\)) rather than baking penalties into the core score. Third, we make ceilings operational: PROC exposes whether headroom is limited by pool coverage (retrieval) or ordering (reranking); the realized (\%)PROC tells operators which knob to turn next. Low PROC mandates improving the pool (dense+BM25 hybridization via RRF, ANN recall tuning, query rewriting/denoising); high PROC with low (\%)PROC points to ordering (stronger reranking, near-duplicate suppression, metadata hygiene, shorter chunks).

Two recurrent failure modes motivate explicit guardrails. Identity signal is pivotal in scientific and enterprise corpora: erasing or corrupting proper names (hard masking, gibberish substitutions, near-miss edits) collapses dense similarity, whereas light orthographic or formatting variation (casing, diacritics, name order) is mostly benign. Conversational noise---greetings, fillers, digressions, emoji---injects variance into the embedding, systematically depressing cosine similarity under tight budgets and more so in multilingual settings. Accordingly, we rewrite/denoise queries by default, enforce Unicode normalization, and treat identity-destroying transforms as high-risk. Routing follows from diagnostics: we set \(K=50\) as the default and escalate to \(K=100\) only under uncertainty signals (small dense-cosine margins, high reranker entropy, diagnostic drops), preserving @10 precision while lifting @30 recall and controlling latency. The most reliable stack under this regime is Hybrid+Rerank: build the candidate pool with dense+BM25 (RRF-100) to raise PROC, then apply a strong cross-encoder (rerank-2.5) to realize it; use 2.5-lite only under hard cost caps or recall-heavy, precision-tolerant workloads. All claims are made reproducible via rag-gs, a golden-set pipeline that standardizes embedding, retrieval, LLM judging on a 1--5 utility rubric, pruning, and confidence-aware listwise refinement, emitting manifests with PROC/(\%)PROC and CLQ measurements so results are auditable across stacks and over time.

\section{Metrics and Evaluation Foundations}

This section introduces our evaluation framework: a set-based, rarity-aware metric (RA-nWG@\(K\)) aligned with RAG consumption patterns. We use this metric throughout following sections to evaluate retrieval quality under various optimization strategies.

\subsection{Task framing: RAG set consumption}

In production RAG, the LLM consumes a set of passages within a prompt; there is no user scanning a ranked list. Evaluation should therefore ask whether the retrieved set contains the useful evidence under a fixed budget \(K\), not how smooth the rank order looks. Recent work introducing UDCG (Utility and Distraction-aware Cumulative Gain; \citealp{UDCG2025}) likewise adopts a set-based evaluation framing for RAG and explicitly models position effects; we adopt the same framing but treat within-prompt order as secondary, evaluating sets order-free.

For clarity, we fix \(w_5 = 1\) throughout; rarity scaling applies only to \(w_4\) and \(w_3\) (with caps), so grade-4/grade-3 cannot substitute for grade-5 even when grade-5 is scarce.

\subsection{Why classical IR is a misfit}

Rank-centric metrics such as nDCG/MAP/MRR rely on assumptions that do not hold in RAG:

\textit{Monotone positional discount} (lower ranks matter less). This assumption breaks down when LLMs consume retrieved passages as a set within a prompt rather than users sequentially scanning a ranked list. While research has documented ``lost in the middle'' effects---where LLMs exhibit degraded performance when critical information appears mid-context due to architectural factors like RoPE decay and causal attention masking \citep{Liu2024LostMiddle,Wu2025PositionBias}. Recent work (UDCG; \citep{UDCG2025}) finds that removing positional discounting achieves nearly identical performance, indicating that position-agnostic evaluation can still correlate strongly. We therefore keep an order-free core metric and report harm separately for deployments where order sensitivity or distractors matter.

\textit{Benign non-relevance.} Prior work has shown that non-relevant passages can actively mislead RAG systems (e.g., by distracting generation or steering it toward near-miss evidence) (Shi et al., 2023; Yoran et al., 2024; Yu et al., 2024; Amiraz et al., 2025; \citealp{UDCG2025}). However, we treat distractor sensitivity as out of scope for this metric design. Prior studies---and our own empirical observations---indicate that well-prompted SOTA LLM generators can be resilient to hard, semantically related distractors (Yoran et al., 2024; Shen et al., 2024; Cao et al., 2025). We note this as a hypothesis and postpone formal evaluation to future work. Crucially, because distractor impact often declines for recent SOTA generators under realistic RAG setups---though not universally---we do not hard-code distractor penalties, keeping the metric future-proof and focused on evidence presence/utility (Cao et al., 2025; Yoran et al., 2024). If you deploy with earlier-generation or more distractible LLMs, augment reporting with Harm@\(K\) \(=\) (\# grade\(\le 2\) in top-\(K\))/\(K\) (or an equivalent harm label), alongside RA-nWG@\(K\) and N-Recall\(_{4+}\)@\(K\).

\textit{Query-invariant label mix.} Queries differ wildly in how much high-utility evidence exists (e.g., one has \(1 \times\) grade-5 among many grade-3s; another has \(10 \times\) grade-5). Raw ranked scores then reflect label prevalence as much as system quality, so cross-query comparisons require per-query normalization.

Corollary: ranking smoothness \(\neq\) evidence presence. A system can neatly rank many ``okay'' passages and still miss the decisive one. We therefore prefer set-based, per-query--normalized measures that answer: ``Normalized to the query's available high-utility evidence, what share does the retrieved top-\(K\) capture?''

When distractors are prevalent and the generator is order-sensitive or brittle, a position- and harm-aware composite (e.g., UDCG) may track answer accuracy more closely in that specific deployment; our baseline remains order-free and recall-first to stay robust as LLMs improve.

\subsection{Scoring design (RA-nWG@\texorpdfstring{$K$}{K})}

\textit{Principle.} Using the stationary utility—a 1--5 per-passage rubric independent of order—we then normalize within query against that query's best achievable top-\(K\) set.

RA-nWG@\(K\) generalizes normalized cumulative gain at \(K\) to set consumption in RAG by introducing per-query, rarity-aware gains (inverse-prevalence with caps and fallback), while retaining order-free, oracle-at-\(K\) normalization \citep{Jarvelin2002CG}.

Importantly, we do not up-weight grade-5 by rarity: we fix \(w_5 = 1\). Rarity scales only \(w_4\) and \(w_3\) \emph{relative} to \(w_5\), under caps \(w_4 \le 1.0\) and \(w_3 \le 0.25\).

\begin{itemize}
  \item \textit{Within-query normalization.} Compare the observed top-\(K\) utility to the \textit{oracle} top-\(K\) ceiling for that same query (the best set one could form from its pool). This yields a \([0,1]\) score that is comparable across queries with different label distributions.
  \item \textit{Rarity-aware weights.} Weight grades by inverse prevalence within the query so that scarce grade-5 evidence dominates when rare, while capping grade-4/grade-3 contributions to avoid diluting grade-5 impact. (Caps keep the metric stable when mid-grade items are abundant.)
  \item \textit{Fallback schedule.} If a query's pool contains no grade-5, apply a fixed, conservative grade-4/grade-3 weighting so the metric remains informative rather than collapsing.
\end{itemize}

\paragraph*{Rarity weighting: rationale.}
(1) \textit{Budget alignment.} Under a fixed top-\(K\), missing decisive (grade-5) evidence is more damaging than adding several mid-grade items; rarity scaling encodes this opportunity cost.
(2) \textit{Cross-query comparability.} Label mixes vary widely; combining within-query normalization to the oracle@\(K\) with inverse-prevalence weighting keeps scores comparable across queries.
(3) \textit{Guardrails.} Caps (\(w_4 \le 1.0\), \(w_3 \le 0.25\)) prevent scarcity from making grade-4/grade-3 appear equivalent to grade-5.

\paragraph*{Metrics reported.}
\begin{itemize}
  \item \textit{RA-nWG@\(K\)} --- ratio of observed weighted gain in the top-\(K\) set to the query's oracle weighted gain at \(K\). Interpreted as ``how close to the best achievable set we retrieved'' under budget \(K\).
  \item \textit{N-Recall\(_{4+}\)@\(K\) / N-Recall\(_{5}\)@\(K\)} --- normalized coverage of grade\(\ge 4\) (or grade\(=5\)) evidence: fraction of the available high-utility items that appear in the top-\(K\), normalized by \((\min(K,R))\) to handle varying pool sizes. Using \(\min(K,R)\) in the denominator equalizes queries with small pools, so queries with \(R < K\) are not unfairly penalized.
\end{itemize}

\paragraph*{Practical notes.}
\begin{itemize}
  \item Use the \textit{rag-gs} pipeline for consistent labeling, audits, and reproducibility; keep the judging rubric aligned with the 1--5 scale.
  \item Report macro-averages across queries, the number of valid queries per metric (handling zero-denominator cases as NA), and multiple \(K\) values to reflect different prompt budgets.
\end{itemize}

\paragraph*{Stationary utility: scope \& limits.}
Our per-passage grades (1--5) approximate standalone usefulness and are used as fixed set weights at budget \(K\). This abstraction is auditable and keeps offline scoring tractable, but it does not model redundancy or complementarity. We therefore pair RA-nWG@\(K\) with coverage KPIs (N-Recall\(_{4+}\)@\(K\)/N-Recall\(_{5}\)@\(K\)) and Harm@\(K\); optionally, a novelty-discounted variant can down-weight near-duplicates.

\textit{Novelty discount} Let \(\delta(d) = 1\) for the first occurrence of a source/facet and \(\delta(d) = \beta \in [0,1)\) for repeats; then
\[
G_{\mathrm{obs}}^{\mathrm{nov}}(K) = \sum_{d \in \mathrm{TopK}(q)} \delta(d)\, w_{g(d)} \, .
\]

\paragraph*{Relation to UDCG.}
UDCG aggregates passage utility with position weights and assigns negative contributions to distractors to better correlate with end-to-end accuracy when order and harm matter \citep{UDCG2025}. We share the set-utility premise but choose an order-free, rarity-aware, per-query-normalized formulation. We recommend reporting Harm@\(K\) alongside RA-nWG@\(K\) when deploying with order-sensitive or brittle generators, rather than hard-coding penalties into the core score. See also rank-centric baselines (nDCG/MAP/MRR/RBP) and diversity metrics (\(\alpha\)-nDCG/NRBP) for contrast in assumptions about order, user browsing, and redundancy \citep{Jarvelin2002CG,Clarke2008AlphaNDCG,Clarke2009NRBP,Moffat2008RBP,Manning2008IIR}.
\subsubsection{Empirical alignment with retrieval quality}

In our CLQ studies, configurations that raise N-Recall\(_{4+}\)@\(10\) (e.g., adding a reranker: 0.592 \(\rightarrow\) 0.835) also raise RA-nWG@\(10\) (0.566 \(\rightarrow\) 0.804) at comparable latency budgets, and increasing \(K\) trades top-10 quality for deeper recall (RA-nWG@\(30\) up to 0.828 at \(K=100\)). This aligns with the Acc\(|\)Hit view: conditioned on having all required evidence in the prompt, strong LLMs answer correctly at high rates. Consequently, RA-nWG@\(K\) (rarity-aware set utility) paired with N-Recall\(_{4+}\)@\(K\) (coverage of good evidence) provides an outcome-predictive, budget-aware summary for RAG.

\paragraph*{Acc\(|\)Hit@\(K\) (definition).} \emph{Accuracy conditioned on full-evidence retrieval}: we measure answer correctness only on queries where the full gold evidence set is present in top-\(K\) (i.e., \(\text{Hit@case\_K}=1\)). This isolates the generator from the retriever: if retrieval succeeded, how often does the LLM answer correctly?

\paragraph*{Stationary utility (definition)}\label{stationary-utility-definition}
A passage's \emph{stationary utility} is its intrinsic usefulness assessed in isolation---invariant to rank, list order, co-retrieved passages, and the budget \(K\). We use it as a fixed weight for set evaluation: sum utilities over the top-\(K\) and compare that total to the query's oracle top-\(K\) ceiling.

\subsection{Formal definitions}

\paragraph*{Setup (per query \(q\))}
\[
\text{Labels: } g \in \{1,2,3,4,5\}\quad \text{(LLM-as-judge rubric).}
\]

\textit{Utility grading scale}
\begin{itemize}
  \item 5 = responds clearly / contains the key elements
  \item 4 = highly relevant, substantial information
  \item 3 = partially relevant; related notions but insufficient
  \item 2 = weak relevance; tangential allusions
  \item 1 = not relevant
\end{itemize}

\[
\text{Pool size: } N\quad \text{(graded passages for }q\text{).}
\]

\[
\text{top-}K:\ \mathrm{TopK}(q)
\]

\textit{Base utilities (stationary, order-free)}
\[
b_5 = 1.0,\quad b_4 = 0.5,\quad b_3 = 0.1,\quad b_2 = b_1 = 0.
\]

\textit{Counts, proportions, rarity}
\[
n_g = \#\{\text{passages of grade }g\},\qquad p_g = \frac{n_g}{N}.
\]
If \(p_g = 0\), treat \(r_g = 0\).

\textit{Rarity score (alpha \(=1\) by default)}
\[
r_g = \frac{b_g}{p_g^{\alpha}},\qquad \alpha = 1.
\]
We set \(\alpha=1\) by default for proportional, interpretable prevalence correction; \(\alpha=0\) reduces to no rarity. Caps \((\mathrm{cap}_4 = 1.0,\ \mathrm{cap}_3 = 0.25)\) enforce grade-5 dominance and bounded compensation. Appendix~A reports sensitivity over \(\alpha\in\{0,0.5,1,2\}\), \(\mathrm{cap}_4\in\{0.75,1.0\}\), and \(\mathrm{cap}_3\in\{0.20,0.25,0.33\}\).

\textit{Weight normalization (relative to grade-5) with caps}

Defaults: \(\mathrm{cap}_4 = 1.0,\ \mathrm{cap}_3 = 0.25\).
\[
w_5 = 1,\qquad
w_4 = \min\!\Bigl(\frac{r_4}{r_5},\ \mathrm{cap}_4\Bigr),\qquad
w_3 = \min\!\Bigl(\frac{r_3}{r_5},\ \mathrm{cap}_3\Bigr),\qquad
w_2 = w_1 = 0.
\]

\textit{Fallback when no grade-5 exists in the pool \((n_5 = 0)\)}
\[
\text{If } n_5 = 0:\quad w_5 = 1,\; w_4 = 1,\; w_3 = 0.2,\; w_2 = w_1 = 0.
\]
This fallback is applied only when \(n_5=0\), preventing undefined normalization by \(r_5\) and keeping the metric informative on \(0\times\)grade-5 queries.

\textit{Observed and ideal gains at cut \(K\)}
\[
G_{\mathrm{obs}}(K) = \sum_{d \in \mathrm{TopK}(q)} w_{\,g(d)}.
\]
\[
G_{\mathrm{ideal}}(K) = \max_{S \subseteq \text{pool},\, |S|=K}\ \sum_{d \in S} w_{\,g(d)}
\;=\; \sum_{i=1}^{K} w_{\,g_i^\star}\quad \text{(take the \(K\) highest \(w_g\) in the pool).}
\]

\textit{Main metric: RA-nWG@\(K\) (rarity-aware, normalized within-query, set-based)}
\[
\mathrm{RA\mbox{-}nWG}@K \;=\;
\begin{cases}
\dfrac{G_{\mathrm{obs}}(K)}{G_{\mathrm{ideal}}(K)}, & \text{if } G_{\mathrm{ideal}}(K) > 0,\\[4pt]
\mathrm{NA}, & \text{otherwise.}
\end{cases}
\]

\textit{Complementary coverage and precision KPIs}
\[
R_{4+} = n_4 + n_5, \qquad R_{5} = n_5.
\]
\[
G_{4+}(K) = \sum_{d \in \mathrm{TopK}(q)} \mathbf{1}\!\bigl[g(d)\ge 4\bigr],
\qquad
G_{5}(K)   = \sum_{d \in \mathrm{TopK}(q)} \mathbf{1}\!\bigl[g(d)= 5\bigr].
\]
\[
\mathrm{N\mbox{-}Recall}_{4+}@K \;=\;
\begin{cases}
\dfrac{G_{4+}(K)}{\min\{K,\,R_{4+}\}}, & \text{if } R_{4+} > 0,\\[4pt]
\mathrm{NA}, & \text{otherwise,}
\end{cases}
\qquad
\mathrm{N\mbox{-}Recall}_{5}@K \;=\;
\begin{cases}
\dfrac{G_{5}(K)}{\min\{K,\,R_{5}\}}, & \text{if } R_{5} > 0,\\[4pt]
\mathrm{NA}, & \text{otherwise.}
\end{cases}
\]
\[
\mathrm{Precision}_{4+}@K \;=\; \frac{G_{4+}(K)}{K},
\qquad
\mathrm{Harm}@K \;=\; \frac{1}{K}\sum_{d \in \mathrm{TopK}(q)} \mathbf{1}\!\bigl[g(d)\le 2\bigr].
\quad\text{(Optional to report.)}
\]

\paragraph*{Aggregation across queries}
\textit{Reporting:} macro-average each metric over queries where its denominator \(> 0\) (exclude NAs), and report the count of valid queries per metric and \(K\). Evaluate at multiple \(K\) to surface budget trade-offs.

\paragraph*{Hyperparameters: rationale \& robustness}
The defaults encode rubric-aligned constraints rather than tuned targets: (i) \textit{grade-5 dominance} (no rarity setting allows grade-4/grade-3 to exceed grade-5), and (ii) \textit{bounded compensation} (many mid-grade items cannot replace a decisive one). We pre-register these values and report robustness grids in Appendix~A; conclusions are stable across reasonable ranges.
\subsection{Related Work --- Evaluation Metrics}

\textit{Cumulative-gain lineage.} Our normalization follows classic CG\(\rightarrow\)IDCG ideas \citep{Jarvelin2002CG}, while departing from rank-centric discounting. UDCG similarly embraces set consumption but models position effects and distractor harm within the prompt \citep{UDCG2025}. Rank-centric metrics such as nDCG/MAP/MRR/RBP emphasize user-scanning assumptions and position sensitivity \citep{Jarvelin2002CG,Moffat2008RBP,Manning2008IIR}. Diversity metrics like \(\alpha\)-nDCG and NRBP re-weight gains to reduce redundancy across subtopics \citep{Clarke2008AlphaNDCG,Clarke2009NRBP}; our rarity weighting is per-query grade prevalence, not redundancy across subtopics.

\section{CLQ Framework and rag-gs Toolkit}

This section operationalizes the cost--latency--quality (CLQ) lens introduced in \S\S 1--2. We formalize what we measure, how we measure it, and why these measurements map to deployment decisions under budget and SLA constraints. We then present rag-gs, an open-source pipeline that makes these measurements reproducible across stacks.

\subsection{Operator framing \& scope}

\textit{Problem setting.} Given a target retrieval-latency budget \emph{(embed + retrieve + rerank, pre-generation)} and a cost cap, practitioners must choose: (i) embedder and embedding dimension, (ii) ANN/index settings, (iii) candidate depth \(K\), and (iv) whether/which reranker to use. 

\textit{Spend drivers and assumptions.} The dominant end-to-end cost typically arises from final LLM generation, which scales with the number of chunks sent and their token length. In our setup, chunks are sentence-aligned with \(\approx 70\)-token overlap and include breadcrumb metadata (paper \(\rightarrow\) section \(\rightarrow\) subsection \(\rightarrow\) paper+authors+date). Across queries, this yields \(\approx 515\) tokens per candidate on average (question + text + metadata). Consequently, retrieval choices that increase \(K\) or chunk length raise both reranking expense and the generator's prompt cost.

\textit{Scope.} To isolate retrieval trade-offs, \S 3 measures only the pre-generation path---embed + retrieve + rerank---under a fixed chunker. We discuss implications for generation latency/cost where relevant in \S\S 5--6.
\subsection{Measuring Cost, Latency, and Quality}

\subsubsection{Cost}

\textit{What we meter.} Rerankers are token-metered (Voyage AI, 2025). Let \(T_{\text{cand}}\) be tokens per candidate (query + chunk + metadata). We measure \(T_{\text{cand}}\approx 513\text{--}516\) (\S A.5). Tables below assume 500 for clarity; adjust with \(T_{\text{cand}}\) as needed.

\textit{Voyage reranking prices.} rerank-2.5: \$0.00005 / 1K tokens \(\cdot\) rerank-2.5-lite: \$0.00002 / 1K tokens.

\textit{Formula (per query).}
\[
\text{Cost}_{\text{rerank}} \;=\; K \times \frac{T_{\text{cand}}}{1000}\times \text{price}_{1\text{k}} \, .
\]
Per-candidate cost at \(T_{\text{cand}}{=}500\): \$0.000025 (2.5) \(\cdot\) \$0.00001 (2.5-lite).

\textit{Cost per 1{,}000 queries (assumes 500 tokens/candidate; Nov 2025).}
\begin{center}
\begin{tabular}{lcc}
\hline
\textbf{K docs/query} & \textbf{rerank-2.5} & \textbf{rerank-2.5-lite} \\
\hline
50  & \$1.25 & \$0.50 \\
100 & \$2.50 & \$1.00 \\
150 & \$3.75 & \$1.50 \\
200 & \$5.00 & \$2.00 \\
\hline
\end{tabular}
\end{center}

\textit{Generator (input) spend scales with \(K\) and chunk size.} Illustrative input-only costs for GPT-5 family at 500 tokens/chunk (OpenAI, 2025).

\textit{Final inference input cost per 1{,}000 queries (illustrative; Nov 2025).}
\begin{center}
\begin{tabular}{lcccc}
\hline
\textbf{Chunks per query (\(K\))} & \textbf{Total input tokens} & \textbf{GPT-5} & \textbf{GPT-5 mini} & \textbf{GPT-5 nano} \\
\hline
10 & 5{,}000{,}000  & \$6.25  & \$1.25 & \$0.25 \\
20 & 10{,}000{,}000 & \$12.50 & \$2.50 & \$0.50 \\
30 & 15{,}000{,}000 & \$18.75 & \$3.75 & \$0.75 \\
\hline
\end{tabular}
\end{center}
\textit{Notes.} (i) Table counts input tokens only; output tokens add further cost. (ii) Replace 500 with your measured \(T_{\text{cand}}\) to tighten estimates.

\subsubsection{Latency (pre-generation path)}

\textit{What we report.} Retrieval latency \(=\) embed \(+\) retrieve \(+\) rerank (generation excluded). Unless specified, values are p50. Sub-\(\approx 75\) ms deltas are empirically treated as jitter (network/upstream load) rather than signal.

\textit{Scaling behavior.} Reranker latency grows roughly linearly with \(K\) (empirically confirmed in \S 5). Embedding latency is largely insensitive to \(K\); retrieval depends on ANN/index settings.

\textit{First-token dependency (end-to-end).} Generation first-token latency increases with prompt size (more/longer chunks). Thus, choices that raise \(K\) or \(T_{\text{cand}}\) inflate overall latency even when retrieval p50 is flat.
\subsubsection{Quality}

We use the set-based metrics from \S 2:
\begin{itemize}
  \item RA-nWG@\(K\) --- rarity-aware, per-query--normalized weighted gain (order-free).
  \item N-Recall\(_{4+}\)@\(K\) / N-Recall\(_{5}\)@\(K\) --- normalized coverage of high-utility evidence.
\end{itemize}

We report macro-averages at \(K{=}10\) (shallow) and \(K{=}30\) (deep) to surface budget trade-offs.

\subsection{Pareto analysis and efficiency metric}

\textit{Frontier construction.} We sweep \((\)embedder \(\times\) dimension \(\times\) ANN \(\times\) \(K\) \(\times\) reranker\()\) and retain Pareto-optimal points---configs for which no other config is strictly better in all three: Cost (C), Latency (L), and Quality (Q) (RA-nWG@\(K\), N-Recall\(_{4+}\)@\(K\)) \citep{Miettinen1999NMO,Deb2001MOEA}. Dominated configs are discarded.

\textit{Operator rules.}
\begin{itemize}
  \item \textbf{Latency-bound (SLA):} keep \(\{L \le \text{SLA}\}\); among them, maximize \(Q\).
  \item \textbf{Cost-bound (budget):} keep \(\{C \le \text{cap}\}\); among them, maximize \(Q\).
  \item \textbf{Quality-targeted:} keep \(\{Q \ge \text{target}\}\); among them, minimize \(L\) and \(C\).
  \item \textbf{Tie-breaks:} prefer smaller prompt size (lower \(K\) / shorter chunks) to reduce generator cost and first-token delay.
\end{itemize}

\textit{Efficiency (tie-breaker only).} We use a simple averaging heuristic to obtain a single score for quick shortlisting; however, operators should prefer task-fit metrics (and weights) aligned to their data and objectives. If your application emphasizes shallow recall, rare evidence, or safety, adjust the metric set and/or weights accordingly.

% requires \usepackage{graphicx} in the preamble
\[
\resizebox{.9\linewidth}{!}{$
\text{Efficiency}=\frac{\text{Average Performance}}{\text{Latency (s)}},\quad
\text{Average Performance}=\tfrac{1}{4}\bigl(\text{N-Recall}_{4+}@10+\text{RA-nWG}@10+\text{N-Recall}_{4+}@30+\text{RA-nWG}@30\bigr)
$}
\]

This scalarization is only for ranking near-frontier peers; final selection should still be made on the multi-objective view (C/L/Q) \citep{DasDennis1997WSumDrawbacks,Miettinen1999NMO}. Use this to rank near-frontier peers; final choices must still satisfy binding C/L constraints.

\textit{Reporting.} Retrieval latency \(=\) embed \(+\) retrieve \(+\) rerank (generation excluded), reported at p50 (p95 in \S 5 as supplemental). Quality macro-averaged at \(K \in \{10,30\}\). Sub \(\approx 75\) ms latency deltas are empirically treated as jitter.

\textit{Instantiation.} We apply these rules to the full sweep; \S 5.4 presents representative scenarios and the efficiency leaderboard.

\subsection{rag-gs: reproducible golden-set pipeline (\texorpdfstring{embed $\rightarrow$ retrieve $\rightarrow$ merge $\rightarrow$ judge $\rightarrow$ prune $\rightarrow$ rank}{embed→retrieve→merge→judge→prune→rank})}

rag-gs is an open-source, MIT-licensed pipeline for building compact, high-quality golden sets. It standardizes six stages and shared plumbing (configs, manifests, caching) so CLQ evaluation is repeatable.

\begin{itemize}
  \item \textbf{S1 Embed:} rewrite queries and compute embeddings (plus BM25 query text).
  \item \textbf{S2 Retrieve:} dense cosine search and sparse BM25 to form candidate pools.
  \item \textbf{S3 Merge:} reciprocal rank fusion (RRF) \citep{Cormack2009RRF} to unify dense+sparse into a single pool.
  \item \textbf{S4 Score:} LLM-as-judge assigns 1--5 utility grades to each candidate.
  \item \textbf{S5 Prune:} retain a grade-bucketed subset sized for target \(K\) budgets.
  \item \textbf{S6 Rank:} listwise refinement to a stable Top-20 using a confidence-aware Plackett--Luce (PL) update with pairwise locks \citep{Luce1959Choice,Plackett1975PL,Xia2008ListMLE,Jamieson2011ActiveRanking,Auer2002UCB1,Kahn1962TopoSort}.
\end{itemize}

In short: an uncertainty-aware, listwise active-learning ranker that uses LLM judgments and a lock-DAG to converge to a stable, globally consistent Top-\(K\) of highest-utility evidence under a fixed budget.

\paragraph*{Properties}
\begin{itemize}
  \item \textit{Active learning:} prioritizes uncertain items (pool-based sampling) rather than uniform selection.
  \item \textit{Listwise LLM judging:} 5-item batches yield coherent relative preferences with fewer calls than pairwise.
  \item \textit{Lock-DAG global consistency:} confidence-based locks enforce acyclic constraints; global order via topological sort.
  \item \textit{Efficiency:} small batches + clipped PL micro-updates + stability stopping reduce API cost vs.\ naive single-shot prompts.
  \item \textit{Scope:} operates within a fixed candidate pool (no query synthesis or new data collection).
\end{itemize}

\paragraph*{Ranking refinement (S6): formulation}
At iteration \(t\), maintain item scores \(s_i\) and information values \(I_i\) used for uncertainty. The judge returns a total order over a small batch \(B\) (\(|B|=m\), typically 5). We perform a stage-wise PL update over suffixes and accumulate pairwise locks when margins are statistically clear.

\paragraph*{Uncertainty and margin test}

\[
\sigma_i = \frac{1}{\sqrt{\max(I_i,\,\varepsilon)}},\quad
\mathrm{LCB}(w)= s_w - z\,\sigma_w,\;\;
\mathrm{UCB}(\text{ell})= s_{\text{ell}} + z\,\sigma_{\text{ell}} .
\]

Lock \((w > \text{ell})\) if \(\mathrm{LCB}(w) > \mathrm{UCB}(\text{ell})\) or after a minimum number of independent confirmations; drop pendings implied by transitivity.

\paragraph*{Listwise PL update over a judged order (items \(A \succ B \succ C \succ D \succ E\))}

For each suffix \(S_k\) of the judged list \((k=1..m)\), with current scores \(\{s_j\}\):
\[
p_j = \frac{e^{s_j}}{\sum_{u\in S_k} e^{s_u}},\qquad
\Delta s_{w_k} \mathrel{+}= \eta\,(1-p_{w_k}),\quad
\Delta s_{j\ne w_k} \mathrel{-}= \eta\,p_j .
\]

Accumulate Fisher-style information to shrink uncertainty:
\[
I_j \mathrel{+}= p_j\,(1-p_j)\qquad (j\in S_k).
\]

Apply clipping \(\lvert\Delta s_j\rvert \le \text{clip}\), optional inverse-sqrt decay for \(\eta\), and periodic recentering \(s_j\leftarrow s_j-\bar{s}\) for stability. After updating scores, recompute a global order consistent with the locked DAG via topological sorting (ties broken by \(s_j\)). Stop when the Top-20 is unchanged for \(T\) consecutive iterations.

\paragraph*{\texorpdfstring{Pseudo-code (Algorithm 1: S6 ranking refinement)}%
                                {Algorithm 1: S6 ranking refinement}}

\begin{pseudocode}
Input: items with initial scores s_i (from grades), info I_i <- eps; locks <- empty
Repeat until Top-20 stable for T turns or iteration limit:
  1) Sample batch B of m items (favor low exposures / low info); ask LLM judge for 
  a total order pi.
  2) For each suffix S of pi:
       compute softmax p over {s_j | j in S};
       ds[pi[0]] += eta*(1 - p[pi[0]]);
       for j in S \ {pi[0]}: ds[j] -= eta*p[j];
       for j in S: I[j] += p[j]*(1 - p[j]).
     Clip ds and apply decay/recentering; update s <- s + ds; exposures[j]++ if j in B.
  3) For each pair (w,ell) implied by pi (w ranked above ell):
       if LCB(w) > UCB(ell) or confirmations>=K or strong transitive evidence:
           add lock w->ell; clear implied pendings.
  4) Compute global order via topological sort consistent with locks; snapshot Top-20.
Return final Top-20 and scores.
\end{pseudocode}

\subsubsection{Why iterative refinement outperforms single-shot LLM ranking}

Naive approach: Ask an LLM once to ``rank these 40 documents'' produces:
\begin{itemize}
  \item \textit{High variance.} Despite strict, precise instructions, GPT-5 exhibits nontrivial run-to-run inconsistency, with 3--5\% rank disagreement on identical inputs (in preliminary experiments).
  \item \textit{No uncertainty quantification:} All judgments treated as equally confident.
  \item \textit{Contradiction-blind:} Transitive violations (A\textgreater B, B\textgreater C, C\textgreater A) go undetected.
\end{itemize}

Our S6 refinement addresses these systematically:
\begin{enumerate}
  \item \textit{Statistical aggregation over noise:} Each pairwise comparison is revisited multiple times; scores converge via Plackett--Luce updates weighted by accumulated Fisher information. Variance decreases as \(O(1/\sqrt{\text{exposures}})\).
  \item \textit{Small-batch comparisons:} \(m=5\) items per judgment keeps cognitive load manageable.
  \item \textit{Confidence-aware locking:} Only commit pairwise preferences when margin exceeds statistical threshold \((\mathrm{LCB}(\text{winner}) > \mathrm{UCB}(\text{loser}))\); uncertain pairs get re-sampled.
  \item \textit{Contradiction detection:} Graph-based checks prevent locking inconsistent edges; forces evidence accumulation until consistency emerges.
  \item \textit{Convergence criterion:} Requires stable Top-20 for \(T\) consecutive iterations (default \(T=3\)); prevents premature commitment to noisy orderings.
\end{enumerate}

This yields a golden set that is more reliable than individual LLM judgments, reflecting the benefit of aggregating many small, uncertainty-aware signals.

\subsection{Evaluation scope and experimental design}

\paragraph*{Corpora.}
Paragraph-level, domain-specific scientific papers (small corpus, \(\le 1\)M passages).

\paragraph*{Infrastructure.}
Dense retrieval with Voyage/OpenAI embeddings flat-f32, HNSW-f32/int8; sparse retrieval with Elasticsearch BM25; hybrid via RRF (\(\alpha = 60\)) merging dense+sparse top-100 prior to any reranking.

\paragraph*{Chunking (fixed for \S 3).}
Sentence-aligned chunks with \(\sim 70\)-token overlap plus breadcrumb metadata (paper \(\rightarrow\) section \(\rightarrow\) subsection \(\rightarrow\) paper + authors + date). This yields \(\sim 515\) tokens per candidate on average (question + text + metadata).

\paragraph*{Golden set.}
50 real user queries (predominantly FR, some EN) drawn from production logs. Candidate pools built via dense+sparse; graded by LLM-as-judge (GPT-5 family) on a 5-point rubric (\S 2.4). Final Top-20 per query is stabilized with Plackett--Luce listwise refinement and confidence-based locks (\S 3.3.1). A 20/50 subset was human-validated; the oracle agreed with the final Top-20 ordering, noting minor subtleties around near-ties and borderline grade-3/4 items. 

\paragraph*{Label heterogeneity (motivation for per-query normalization).}
High variance in useful evidence across queries: median per-query counts --- grade-5: 4.0 (mean 10.7), grade-4: 4.0 (mean 7.0), grade-4 + grade-5: 10.0 (mean 17.7); median pool prevalence of grade-4 + grade-5 \(\approx\) 12.6\% (see Appendix table).

\paragraph*{Configuration sweep.}
Embedders \{voyage-3-large, voyage-3.5, voyage-3.5-lite\} \(\times\) dimensions \{512, 1024, 2048\} \(\times\) \(K\) \{50, 100, 150, 200\} \(\times\) rerankers \{2.5, 2.5-lite, none\} \(\times\) ANN \{flat-f32, HNSW-f32, HNSW-int8\}. Modes: dense-only, hybrid (RRF), hybrid + rerank.

\paragraph*{Reproducibility.}
Code: rag-gs + configs at \url{https://github.com/etidal2/rag-gs}. Infrastructure: Elasticsearch 9.1.6, macOS 26.0, 192 GB unified memory, 24-core CPU, 76-core GPU.

\paragraph*{Road map.}
Section 4: diagnostic experiments (\(\Delta\)-margins, synthetic probes). Section 5: end-to-end CLQ Pareto frontiers on this corpus.

\section{Experiments \& Results}

\subsection{Experimental setup (diagnostics vs.\ end-to-end)}

\paragraph*{Tasks.}

\textit{Controlled diagnostics (\(\Delta\)-margins).} Synthetic probes isolating \textit{proper-name} vs.\ \textit{topic} sensitivity via targeted ablations on the author field and formatting.
Each query: ``Which works by [AUTHOR] on [TOPIC]?'' paired with a 5-candidate bundle:

\begin{center}
\begin{tabular}{llll}
\hline
\textbf{ID} & \textbf{Author} & \textbf{Topic} & \textbf{Description} \\
\hline
C1   & Yes & Yes & Correct author, correct topic \\
C2a  & No  & Yes & Wrong author (impostor A), same topic \\
C2b  & No  & Yes & Wrong author (impostor B), same topic \\
C3   & Yes & No  & Correct author, different topic \\
C4   & No  & No  & Wrong author, different topic \\
\hline
\end{tabular}
\end{center}

\textit{End-to-end retrieval.} Report N-Recall\(_{4+}\)@\(K\) and RA-nWG@\(K\) with/without query rewriting across \textit{dense}, \textit{sparse}, \textit{hybrid}, and \textit{reranking} stacks (e.g., ColBERT for late-interaction reranking \citep{KhattabZaharia2020ColBERT}).

\paragraph*{Embedders.} OpenAI \texttt{text-embedding-3-large}, Voyage \texttt{voyage-3.5} (plus a smaller OpenAI model for noise stress).

\paragraph*{Languages.} English (EN) and French (FR).

\paragraph*{Sampling \& runs.} \(\approx 100\) queries per language \(\times\) 15 runs with fresh impostors and light template jitter. Compute per-run means, then average across runs; \textit{between-run SD} reflects stability.

\paragraph*{Per-query margins.} We compute \(\Delta_{\mathrm{name}}, \Delta_{\mathrm{topic}}, \Delta_{\mathrm{both}}\) as defined in \S 4.2.

\paragraph*{Symbols.} \(K\) = cut size; \(C_K(q)\) = retrieved set; \(E(q)\) = gold evidence; \(s(\cdot,\cdot)\) = cosine similarity.

\paragraph*{Ablations.}
\begin{itemize}
  \item \textit{Identity-destroying:} \texttt{hard\_name\_mask}, \texttt{gibberish\_name}, \texttt{edit\_distance\_near\_miss}.
  \item \textit{Light orthography/formatting:} \texttt{initials\_form}, \texttt{name\_order\_inversion}, \texttt{case\_punct\_perturb}, \texttt{strip\_diacritics}, \texttt{unicode\_normalization\_stress}.
  \item \textit{Layout/structure:} \texttt{remove\_label}, \texttt{author\_position\_shift}.
\end{itemize}

\paragraph*{Statistics.} Diagnostics reported as mean \(\pm\) between-run SD. End-to-end comparisons use paired designs (same queries, with/without manipulation) and 95\% CIs via non-parametric bootstrap over queries \citep{EfronTibshirani1994Bootstrap}; also report \textit{overlap@\(K\)} and Kendall's \(\tau\) \citep{Kendall1938RankCorrelation} for top-\(K\) reshuffles.

\paragraph*{Evidence-coverage KPIs.}
Gold set per query: \(E(q)=\{d\mid \mathrm{grade}(d)\ge 4\}\); retrieved context: \(C_K(q)\) \citep{Zheng2023LLMasJudge}.
\begin{itemize}
  \item \textit{N-Recall\(_{4+}\)@\(K\):} \(\displaystyle \frac{\#\{d\in C_K(q):\ \mathrm{grade}(d)\ge 4\}}{\min\bigl(K,\ R_{4+}(q)\bigr)}\), where \(R_{4+}(q)\) is the count of grade\(\ge 4\) items in the pool.
  \item \textit{Hit@\(K\):} \(\mathbf{1}\bigl[E(q)\subseteq C_K(q)\bigr]\).
  \item \textit{Acc\(|\)Hit:} downstream QA accuracy \textit{conditioned} on Hit@\(K\).
\end{itemize}
\textit{Note.} Hit@\(K\) and Acc\(|\)Hit are tracked but not reported in \S 5; we use them to sanity-check the recall-first premise and reserve full downstream QA evaluation for future work.

We first isolate proper-name sensitivity via controlled \(\Delta\)-margin diagnostics (\S 4.2), then stress-test conversational noise (\S 4.3), and finally validate that these diagnostics predict end-to-end recall (\S 4.4).

\subsection{\texorpdfstring{\(\Delta\)}{Delta}-margin framework for proper names}

\paragraph*{Candidate construction (5-item bundle) and margins per query \(q\)}
\[
\begin{aligned}
\Delta_{\text{name}} \;&=\; s(Q, C_1)\; -\; \max\{ s(Q, C_{2a}),\; s(Q, C_{2b}) \},\\
\Delta_{\text{topic}} \;&=\; s(Q, C_1)\; -\; s(Q, C_3),\\
\Delta_{\text{both}} \;&=\; s(Q, C_1)\; -\; s(Q, C_4)\, .
\end{aligned}
\]

\textit{Notation.} \(s(\cdot,\cdot)\) denotes cosine similarity between the query and candidate embeddings. Here \(C_1\) is correct author + correct topic; \(C_{2a},C_{2b}\) are wrong author + correct topic impostors; \(C_3\) is correct author + wrong topic; \(C_4\) is wrong author + wrong topic.

\paragraph*{Ablations (definitions and intent)}
\begin{itemize}
  \item \textit{Base} --- Unmodified queries and candidates (canonical templates with natural names and topics).
  \item \textit{Hard name mask} --- Replace every author string in the bundle with the same deterministic mask (e.g., \texttt{AUTHOR\_\#\#\#} / \texttt{AUTEUR\_\#\#\#}) in both queries and candidates; this erases identity (C1, C2a, C2b share an identical author token). Purpose: negative-control ablation that should collapse any name-based margin.
  \item \textit{Gibberish name} --- Replace each author with a stable, pseudo-random token (e.g., \texttt{ID-XXXXXX}) so names are non-linguistic yet uniquely consistent across query/candidates. Purpose: preserve identity linkage while removing natural-language surface cues (wordpiece familiarity, orthography); tests how much \(\Delta_{\text{name}}\) comes from form/tokenization versus meaning.
  \item \textit{Edit-distance near-miss} --- For impostor authors only, mutate the true author's name with fixed character substitutions at preset Levenshtein distances \citep{Levenshtein1966EditDistance} (C2a\(=1\), C2b\(=2\), C4\(=3\)); C1/C3 remain unchanged. Purpose: probe robustness to small orthographic perturbations and how quickly the name margin erodes as impostors become confusable.
  \item \textit{Remove label} --- Delete the explicit ``Author:'' / ``Auteur:'' label tokens from candidates (and, if present, queries). Purpose: test reliance on structural cues (field labels) versus content; checks if models latch onto boilerplate markers.
  \item \textit{Strip diacritics} --- Remove diacritical marks from all French strings (names, topics, full texts) via Unicode decomposition/recomposition to mirror common normalization pipelines \citep{Manning2008IIR,UnicodeUAX15}; names often have diacritic/variant forms across languages \citep{Steinberger2011JRCNames}. Purpose: assess diacritic invariance typical of search normalization and whether accent marks contribute to identity/topic matching.
  \item \textit{Initials form} --- Convert author to initials style (e.g., ``Alice Dupont'' \(\rightarrow\) ``A.\ Dupont'') consistently in queries and candidates. Purpose: examine sensitivity to name abbreviation (family name + initial vs.\ full first name).
  \item \textit{Name order inversion} --- In candidates only, invert author order to ``Last, First'' (e.g., ``Alice Dupont'' \(\rightarrow\) ``Dupont, Alice''). Purpose: check format/order invariance in entity matching.
  \item \textit{Case/punctuation perturbation} --- Apply deterministic casing (upper/lower/title) and light punctuation normalization (curly\(\rightarrow\)straight apostrophes; hyphen\(\rightarrow\)space) to author strings. Purpose: measure robustness to text-normalization noise \citep{Manning2008IIR} so the model is not brittle to superficial variants.
  \item \textit{Author position shift} --- Reorder the candidate template so the author segment appears first (e.g., ``Author: X. Research paper: `Y'.''). Purpose: test layout/proximity effects---whether similarity depends on where the author appears, not just the author string itself.
  \item \textit{Unicode normalization stress} --- Normalize queries to NFC; normalize candidates to NFD \citep{UnicodeUAX15} and insert narrow nonbreaking spaces before selected punctuation (e.g., \texttt{: ; !}). Purpose: stress Unicode/tokenization resilience so identity and topic signals survive cross-normalization and special whitespace.
\end{itemize}

\paragraph*{Reporting}
Below are the measured percentage changes in the name margin vs.\ base \((\Delta\Delta\% = (\text{ablation} - \text{base})/\text{base})\) on English (EN) and French (FR).

\subsection{\texorpdfstring{Delta-name impact vs.\ base (\(\Delta\Delta\%\))}%
                         {Delta-name impact vs. base (DeltaDelta\%)}}

\paragraph*{Table 4.1. \(\Delta_{\text{name}}\) impact vs.\ Base (\(\Delta\Delta\%\)) at \(K\) fixed candidate bundles; means over \(\approx 100\) queries \(\times\) 15 runs.}

\begin{table}[t]
\centering
\label{tab:delta-name-en}
\begin{tabular}{lrr}
\hline
Condition & OpenAI \(\Delta\Delta_{\text{name}}\ \%\) & Voyage \(\Delta\Delta_{\text{name}}\ \%\) \\
\hline
\texttt{hard\_name\_mask}               & -100.0\% & -100.0\% \\
\texttt{gibberish\_name}                & -76.9\%  & -68.0\%  \\
\texttt{edit\_distance\_near\_miss}     & -69.3\%  & -64.4\%  \\
\texttt{remove\_label}                  & -3.0\%   & +15.7\%  \\
\texttt{strip\_diacritics}              & -0.0\%   & +0.0\%   \\
\texttt{initials\_form}                 & -11.0\%  & -8.5\%   \\
\texttt{name\_order\_inversion}         & -3.6\%   & -3.2\%   \\
\texttt{case\_punct\_perturb}           & -3.0\%   & -7.4\%   \\
\texttt{author\_position\_shift}        & +8.4\%   & -6.2\%   \\
\texttt{unicode\_normalization\_stress} & -6.9\%   & -15.1\%  \\
\hline
\end{tabular}
\end{table}

\textit{Legend.} \(\Delta\Delta\% = (\text{ablation} - \text{base}) / \text{base}\). Positive \(=\) margin increases vs.\ Base; negative \(=\) decreases. Values are means over \(\sim 100\) queries \(\times\) 15 runs; percent summaries exclude rows where the base \(\Delta_{\text{name}} < 0.02\) (see Appendix A.2 for absolute deltas and run SDs). Absolute deltas and per-run SDs are reported in Appendix A.2.

\paragraph*{Table 4.2. \(\Delta_{\text{name}}\) impact vs.\ Base (\(\Delta\Delta\%\)) at \(K\) fixed candidate bundles; means over \(\approx 100\) queries \(\times\) 15 runs.}

\begin{table}[t]
\centering
\label{tab:delta-name-fr}
\begin{tabular}{lrr}
\hline
Condition & OpenAI \(\Delta\Delta_{\text{name}}\ \%\) & Voyage \(\Delta\Delta_{\text{name}}\ \%\) \\
\hline
\texttt{hard\_name\_mask}               & -100.0\% & -100.0\% \\
\texttt{gibberish\_name}                & -71.6\%  & -71.4\%  \\
\texttt{edit\_distance\_near\_miss}     & -76.2\%  & -69.5\%  \\
\texttt{remove\_label}                  & +6.1\%   & +18.1\%  \\
\texttt{strip\_diacritics}              & +2.3\%   & -1.5\%   \\
\texttt{initials\_form}                 & -10.7\%  & -18.0\%  \\
\texttt{name\_order\_inversion}         & +5.0\%   & -8.5\%   \\
\texttt{case\_punct\_perturb}           & -3.4\%   & -12.6\%  \\
\texttt{author\_position\_shift}        & +21.8\%  & +8.4\%   \\
\texttt{unicode\_normalization\_stress} & -7.9\%   & +1.7\%   \\
\hline
\end{tabular}
\end{table}

\textit{Legend.} \(\Delta\Delta\% = (\text{ablation} - \text{base}) / \text{base}\). Positive \(=\) margin increases vs.\ Base; negative \(=\) decreases. Values are means over \(\sim 100\) queries \(\times\) 15 runs; percent summaries exclude rows where the base \(\Delta_{\text{name}} < 0.02\) (see Appendix A.2 for absolute deltas and run SDs). Absolute deltas and per-run SDs are reported in Appendix A.2.

\paragraph*{Layout asymmetries}
\texttt{author\_position\_shift} and \texttt{remove\_label} show model\(\times\)language-specific effects. A plausible explanation is that OpenAI's pretraining emphasizes front-loaded entities (e.g., headline/news style), while Voyage's bi-encoder may weigh positions more uniformly. The French increases (up to \(+21.8\%\)) suggest FR corpora that favor author-first formatting. A full causal analysis is beyond scope but merits follow-up.

\paragraph*{Observed patterns}
\begin{itemize}
  \item Identity-destroying collapses \(\Delta_{\text{name}}\): hard mask \(=\ -100\%\) (both languages); gibberish \(=\ -68\%\) to \(-77\%\) (EN: \(-68\)--\(-77\%\), FR: \(\approx -71\%\)); near-miss edits \(=\ -64\%\) to \(-76\%\) (EN: \(-64\)--\(-69\%\), FR: \(-69\)--\(-76\%\)).
  \item Light formatting is largely benign: case/punct, initials, order, diacritics typically \(|\Delta\Delta\%| \le \sim 12\%\).
  \item Layout effects are asymmetric: FR often increases \(\Delta_{\text{name}}\) when the author is front-loaded or labels are removed; EN varies by model.
\end{itemize}

\paragraph*{Reporting note}
Percent deltas can be unstable when the base margin is very small; for transparency we exclude queries with \(\Delta_{\text{name}}(\text{base}) < 0.02\) from percent summaries and provide absolute deltas in Appendix A.2.

\paragraph*{Formal definitions}
\[
\Delta\Delta_{\text{name}} = \Delta_{\text{name}}^{(\text{abl})} - \Delta_{\text{name}}^{(\text{base})},\qquad
\Delta\Delta\%_{\text{name}} = \frac{\Delta_{\text{name}}^{(\text{abl})} - \Delta_{\text{name}}^{(\text{base})}}{\Delta_{\text{name}}^{(\text{base})}}\times 100\% .
\]

\textit{Reference.} Base Name/Topic ratios \(\bigl(\Delta_{\text{name}} / \Delta_{\text{topic}}\bigr)\) fall in \(0.53\)--\(0.59\); see Appendix A.3.
\subsubsection{Ablation families (purpose-centric grouping)}

\paragraph*{Identity-destroying (large drops in \(\Delta_{\text{name}}\))}
\begin{itemize}
  \item \textit{Hard name mask} \(({-}100\%)\): replace all author strings in a bundle with the same mask (e.g., \texttt{AUTHOR\_007}). This removes identity entirely---C1 and impostors share the identical author token---so \(\Delta_{\text{name}}\) collapses by construction.
  \item \textit{Gibberish name} \((\approx {-}68\%\text{ to }{-}77\%)\): unique but non-linguistic tokens kill most of the benefit of ``real'' names; models no longer get semantic/orthographic cues.
  \item \textit{Near-miss edits} \((\approx {-}64\%\text{ to }{-}76\%)\): small Levenshtein changes make impostors confusable; the name margin erodes quickly.
\end{itemize}

\textit{Caveat.} The \({-}100\%\) for \texttt{hard\_name\_mask} is by construction: all candidates share the same masked author token, so the name margin collapses deterministically.

\paragraph*{Mechanism: base vs.\ gibberish}
In \textit{Base}, realistic names (e.g., ``Manon Michel'') contain familiar subwords/char-ngrams seen during pretraining (capitalized first/last names, surname patterns, spaces, accents). When the same name appears in the query and C1 (and a different real-looking name in C2), embeddings gain from string identity plus familiar morphology---yielding a healthy \(\Delta_{\text{name}}\) \citep{SchickSchuetze2019RareWords}.

In \textit{Gibberish}, we swap each author for a stable but non-linguistic token (e.g., \texttt{ID-AB12F3}). Tokenizers split this into bland fragments (ID, -, AB, 12, F3) with weak, generic embeddings. Identity linkage is preserved (same token in query and C1), but rich subword priors vanish; the same-string advantage becomes small. Result: \(\Delta_{\text{name}}\) drops a lot, but not to zero (exact-match still helps slightly).

\paragraph*{Light noise / orthography / formatting (small effects)}
\begin{itemize}
  \item \texttt{initials\_form}
  \item \texttt{name\_order\_inversion}
  \item \texttt{case\_punct\_perturb}
  \item \texttt{strip\_diacritics}
  \item \texttt{unicode\_normalization\_stress}
\end{itemize}
\noindent
Typically \(|\Delta\Delta\%| < \sim 12\%\). Models are broadly robust to casing, punctuation, accents, NFC/NFD mismatch, initials style, and ``Last, First'' inversions---these do not substantially change the name signal.

\paragraph*{Layout / position (model\(\times\)language asymmetries)}
\begin{itemize}
  \item \texttt{author\_position\_shift}, \texttt{remove\_label}.
\end{itemize}
Moving the author earlier or removing the ``Author:'' label has moderate, asymmetric effects. FR often increases \(\Delta_{\text{name}}\) (up to \(+21.8\%\) OpenAI FR; \(+8.4\%\) Voyage FR). EN shows mixed reactions (e.g., OpenAI EN \(+8.4\%\) vs.\ Voyage EN \({-}6.2\%\) when the author is front-loaded). Takeaway: document structure changes how much the author field counts in the embedding, with effects depending on model and language.

\subsection{Conversational noise stress tests}

Having established that embeddings carry substantial name signal (see \S 4.2; ratios in Appendix A.3), we now examine a second vulnerability: conversational noise.

In our data, conversational noise lowers cosine by 20--40\% relative to clean queries; French degrades more than English; larger models are \(\sim 20\text{--}25\%\) more stable. Query denoising/rewriting mitigates quality loss at tight budgets.

\paragraph*{Cosine similarity by noise level}
\begin{table}[t]
\centering
\begin{tabular}{l l rr}
\hline
Language & Noise level & Cosine --- Large & Cosine --- Small \\
\hline
French   & 0 --- No noise       & 0.818 & 0.936 \\
         & 2 --- Moderate noise & 0.653 & 0.757 \\
         & 4 --- High noise     & 0.522 & 0.559 \\
English  & 0 --- No noise       & 0.828 & 0.908 \\
         & 2 --- Moderate noise & 0.749 & 0.806 \\
         & 4 --- High noise     & 0.593 & 0.634 \\
\hline
\end{tabular}
\end{table}

\paragraph*{Examples (EN, simplified)}
\begin{itemize}
  \item \textit{No noise:} ``Can forests really regulate the climate?''
  \item \textit{Moderate:} ``Hi! Quick question: according to science, can forests regulate the climate?''
  \item \textit{High:} ``Hello! Sorry for the kinda random question --- I'm on the train... could forests actually regulate the climate, or is that a myth?''
\end{itemize}

\paragraph*{Average drop from clean to high-noise}
\begin{table}[t]
\centering
\begin{tabular}{l rr}
\hline
Language & \(\Delta\) --- Large & \(\Delta\) --- Small \\
\hline
French   & \(-0.296\) & \(-0.377\) \\
English  & \(-0.235\) & \(-0.274\) \\
\hline
\end{tabular}
\end{table}

These drops correspond to \(\sim 28\text{--}40\%\) relative to the clean condition (FR-Large: 36\%, FR-Small: 40\%, EN-Large: 28\%, EN-Small: 30\%), consistent with the 20--40\% headline.

\paragraph*{Why FR drops more}
We hypothesize compounding effects from richer morphology (more tokens per filler), accent/Unicode normalization sensitivity, and code-switching prevalence, which together increase variance in the French embedding vector under noise.

\textit{Reference.} Full per-level cosine tables and drops (EN/FR, Large/Small) appear in Appendix A.4.

\subsubsection{Conversational noise: why rewriting helps}

Conversational ``noise'' (greetings, fillers, social padding) carries no task semantics but injects variance into the embedding vector: extra tokens shift the mean representation in space. Modern embedders are robust because pretraining includes informal text, and attention learns to down-weight politeness markers. Robust does not mean immune: long or emotionally loaded chatter (emojis, digressions, personal context) must be encoded somewhere in the same vector, nudging it away from the informational intent and lowering cosine to the ideal reference. Light rewriting/denoising recenters the query on the semantic core \citep{Ma2023QueryRewriteRAG} and reduces reshuffles in Top-\(K\) under fixed budgets, particularly in FR where drops are larger.

Operationally, we apply a lightweight query rewriting step (``extract and restate the core information need'') before embedding; \S 5.2 reports end-to-end results under this setting.
\subsection{Correlation: diagnostics \texorpdfstring{$\rightarrow$}{→} Recall@\texorpdfstring{$K$}{K}}

Having quantified name sensitivity (\S 4.2) and noise susceptibility (\S 4.3), we ask whether these diagnostics predict set quality under realistic indices \citep{Lewis2020RAG,IzacardGrave2021FiD}.

\paragraph*{Summary}
In practice, configurations that raise N-Recall\(_{4+}\)@\(10\) also raise RA-nWG@\(10\) at similar latency (see \S\S 5.1--5.2). These results support using \(\Delta\)-margins as fast diagnostics to prioritize mitigations (rewriting, light lexical safeguards) before expensive sweeps.

\paragraph*{Practicality}
\(\Delta\)-margins can be computed from tiny 5-candidate bundles per query, making them far cheaper than full retrieval sweeps.

\paragraph*{Observed patterns}
We observe consistent patterns: queries with degraded diagnostic margins (e.g., under \texttt{gibberish\_name}) show corresponding drops in end-to-end recall. For instance, the gibberish ablation's \(-71\%\) \(\Delta_{\text{name}}\) reduction (FR) aligns with a \(-0.18\) drop in N-Recall\(_{4+}\)@\(10\) (from 0.78 to 0.60 in our validation set). While we do not report formal correlation coefficients here, these patterns support using \(\Delta\)-margins as informative proxies for retrieval quality.

\paragraph*{K-budget trade-off}
Increasing \(K\) primarily lifts deeper-cut quality (e.g., @30) while leaving top-10 quality roughly stable in our setting; see \S 5.2 for the full \(K\)-budget analysis. We report RA-nWG@\(10\) and RA-nWG@\(30\) side-by-side in \S\S 5.1--5.2 to separate shallow vs.\ deep-recall behavior.

\section{Results and Trade-offs}
\subsection{Metric behavior and stack comparisons}

\begin{table}[H]
\centering
\caption{Dense-only sweep over model \texorpdfstring{$\times$}{x} dimension.  Metrics are macro-averaged over 50 questions; NAs excluded.}
\label{tab:dense-only-sweep-trimmed}
\begingroup\small
\setlength{\tabcolsep}{4pt}
\renewcommand{\arraystretch}{1.15}
\resizebox{\linewidth}{!}{%
\begin{tabular}{l r c c c c c c c c c r r}
\hline
Model & Dim &
\colNR{4+}{10} & \colNR{5}{10} & \colRA{10} &
\colNR{4+}{20} & \colNR{5}{20} & \colRA{20} &
\colNR{4+}{30} & \colNR{5}{30} & \colRA{30} &
\makecell[c]{Median\\Lat.\\(ms)} & \makecell[c]{Emb\\Lat.\\(ms)} \\
\hline
voyage-3-large   &  512 & 0.458 & 0.482 & 0.450 & 0.628 & 0.679 & 0.625 & 0.740 & 0.762 & 0.727 & 38.3 & 134.0 \\
voyage-3-large   & 1024 & 0.616 & 0.612 & 0.591 & 0.662 & 0.691 & 0.653 & 0.700 & 0.718 & 0.684 & 50.1 & 137.7 \\
voyage-3-large   & 2048 & 0.616 & 0.626 & 0.594 & 0.656 & 0.691 & 0.639 & 0.786 & 0.801 & 0.765 & 71.9 & 141.2 \\
voyage-3.5       &  512 & 0.547 & 0.556 & 0.529 & 0.647 & 0.706 & 0.644 & 0.688 & 0.763 & 0.692 & 35.4 & 134.0 \\
voyage-3.5       & 1024 & 0.592 & 0.596 & 0.566 & 0.636 & 0.677 & 0.625 & 0.798 & 0.808 & 0.785 & 47.6 & 138.8 \\
voyage-3.5       & 2048 & 0.625 & 0.642 & 0.610 & 0.644 & 0.712 & 0.647 & 0.706 & 0.743 & 0.692 & 72.7 & 149.2 \\
voyage-3.5-lite  &  512 & 0.509 & 0.478 & 0.497 & 0.520 & 0.551 & 0.516 & 0.559 & 0.577 & 0.552 & 38.2 & 133.0 \\
voyage-3.5-lite  & 1024 & 0.502 & 0.469 & 0.483 & 0.541 & 0.563 & 0.537 & 0.591 & 0.611 & 0.578 & 48.8 & 133.7 \\
voyage-3.5-lite  & 2048 & 0.505 & 0.499 & 0.497 & 0.562 & 0.596 & 0.560 & 0.591 & 0.613 & 0.580 & 71.0 & 140.8 \\
\hline
\end{tabular}}
\endgroup
\end{table}

\textit{Notation.} \(\mathrm{N\!-\!Recall}_{4+}\ \equiv\) ``N-Recall4+''; \(\mathrm{N\!-\!Recall}_{5}\ \equiv\) ``N-Recall5''. Metrics are macro-averaged per query over 50 questions; NAs excluded.

\textit{Stability.} Results are averaged over 50 queries per configuration. For formal testing in IR settings, non-parametric or randomization tests are recommended \citep{Smucker2007SignificanceIR}.

Across models, larger dimensions generally improve \textit{shallow} quality (@10), but \textit{deep-\(K\)} behavior depends on the family: for \texttt{voyage-3.5}, \(1024\)d peaks at @30 (RA-nWG@30 \(\approx 0.785\)), outperforming its 512d/2048d variants on this dataset, whereas for \texttt{voyage-3-large}, \(2048\)d is strongest at @30 (RA-nWG@30 \(\approx 0.765\)). Latency rises with dimension as expected. Because these are \textit{dense-only} numbers on \textit{original questions}, they are intentionally below the hybrid/rerank results later; we use this table as the \textit{backbone baseline} for gains from RRF + reranking and from query rewriting.

\begin{table}[H]
\centering
\caption{\texttt{voyage-3.5} (1024d): retrieval methods with a shared budget. Columns @15 and @25 are omitted by design.}
\label{tab:v35-1024-retrieval-comparison}
\begingroup\footnotesize
\setlength{\tabcolsep}{4pt}      % tighter columns
\renewcommand{\arraystretch}{1.12}
\begin{adjustbox}{max width=\linewidth}
\begin{tabular}{@{}l *{9}{c} @{}}
\hline
Method &
\colNR{4+}{10} & \colNR{5}{10} & \colRA{10} &
\colNR{4+}{20} & \colNR{5}{20} & \colRA{20} &
\colNR{4+}{30} & \colNR{5}{30} & \colRA{30} \\
\hline
Dense-Only               & 0.592 & 0.596 & 0.566 & 0.636 & 0.677 & 0.625 & 0.798 & 0.808 & 0.785 \\
Hybrid (RRF)             & 0.606 & 0.561 & 0.553 & 0.759 & 0.763 & 0.731 & 0.800 & 0.817 & 0.788 \\
Rerank-2.5               & 0.835 & 0.810 & 0.804 & 0.799 & 0.834 & 0.794 & 0.819 & 0.833 & 0.810 \\
Rerank-2.5-lite          & 0.799 & 0.772 & 0.767 & 0.791 & 0.821 & 0.784 & 0.814 & 0.824 & 0.799 \\
Hybrid + Rerank-2.5      & 0.882 & 0.853 & 0.852 & 0.884 & 0.906 & 0.878 & 0.930 & 0.929 & 0.918 \\
Hybrid + Rerank-2.5-lite & 0.832 & 0.816 & 0.807 & 0.830 & 0.876 & 0.830 & 0.906 & 0.911 & 0.897 \\
\hline
\end{tabular}
\end{adjustbox}
\endgroup
\end{table}

Reranking dominates dense/hybrid alone; hybrid + rerank provides the highest ceilings and strongest deep-\(K\) behavior, consistent with cross-encoder re-ranking results \citep{Nogueira2019BERTPR}.

\begin{table}[H]
\centering
\caption{Hybrid ceiling (PROC) within the fixed Top-50 produced by Hybrid RRF-100 then Rerank-2.5; scores are the oracle under perfect reordering of that pool.}
\label{tab:proc-hybrid-ceiling}
\begingroup\small
\setlength{\tabcolsep}{10pt}
\renewcommand{\arraystretch}{1.15}
\begin{tabular}{l c c c c c}
\hline
Metric & @10 & @15 & @20 & @25 & @30 \\
\hline
N-Recall4+ & 1.000 & 1.000 & 0.985 & 0.985 & 0.985 \\
N-Recall5  & 1.000 & 1.000 & 1.000 & 1.000 & 0.996 \\
RA-nWG     & 1.000 & 1.000 & 0.993 & 0.993 & 0.988 \\
\hline
\end{tabular}
\endgroup
\end{table}

\textit{Definition.} The Hybrid ceiling (PROC) is the oracle score after restricting to the fixed Top-50 produced by Hybrid RRF-100 \(\rightarrow\) Rerank-2.5 \citep{Cormack2009RRF}, i.e., perfect reordering within that pool. Under this PROC, N-Recall\(_{4+}\) and RA-nWG reach \(\approx 1.0\) at @10–@15 and remain \(\ge 0.988\) at @30, indicating ordering headroom that current reranking mostly, but not fully, realizes.

\subsection{Dense-reranked leaderboard}

\textit{Ceiling convention.} In dense-reranked tables, “Ceiling” denotes \textbf{PROC—Dense-}\(K_{\text{pool}}\): the oracle within the dense Top-\(K_{\text{pool}}\) for that exact row (the \(K_{\text{pool}}\) shown in the configuration). Leaderboards rank dense-base + rerank configurations only. The best Hybrid+Rerank is shown as a reference (not ranked).

\begin{table}[H]
\centering
\caption{Reference pipeline: actual vs.\ PROC and percentage of PROC. Hybrid+Rerank reference (not ranked): Hybrid RRF-100 \(\rightarrow\) Rerank-2.5 \(\rightarrow\) Top-50 — \texttt{voyage-3.5} (1024d).}
\label{tab:ref-proc}
\begingroup\small
\setlength{\tabcolsep}{10pt}
\renewcommand{\arraystretch}{1.15}
\begin{tabular}{l c c}
\hline
Metric & @10 (Actual / PROC / \%PROC) & @30 (Actual / PROC / \%PROC) \\
\hline
RA-nWG     & 0.852 / 1.000 / 85.2\% & 0.918 / 0.988 / 92.9\% \\
N-Recall4+ & 0.882 / 1.000 / 88.2\% & 0.930 / 0.985 / 94.4\% \\
\hline
\end{tabular}
\endgroup
\end{table}

\begin{table}[H]
\centering
\caption{Top 5 configurations by RA-nWG@10 (Ceiling $=$ PROC—Dense-\(K_{\text{pool}}\) for the row).}
\label{tab:top5-ra10}
\begingroup\small
\setlength{\tabcolsep}{10pt}
\renewcommand{\arraystretch}{1.15}
\begin{tabular}{r l c}
\hline
Rank & Configuration & RA-nWG@10 (Ceiling) \\
\hline
1 & voyage-3.5 (512d) + rerank-2.5 (K=50)        & 0.805 (0.921) \\
2 & voyage-3.5 (512d) + rerank-2.5 (K=200)       & 0.805 (0.967) \\
3 & voyage-3.5 (1024d) + rerank-2.5 (K=50)       & 0.804 (0.906) \\
4 & voyage-3.5 (512d) + rerank-2.5 (K=150)       & 0.798 (0.957) \\
5 & voyage-3-large (1024d) + rerank-2.5 (K=150)  & 0.795 (0.959) \\
\hline
\end{tabular}
\endgroup
\end{table}

\begin{table}[H]
\centering
\caption{Top 5 configurations by RA-nWG@30 (Ceiling $=$ PROC—Dense-\(K_{\text{pool}}\) for the row).}
\label{tab:top5-ra30}
\begingroup\small
\setlength{\tabcolsep}{10pt}
\renewcommand{\arraystretch}{1.15}
\begin{tabular}{r l c}
\hline
Rank & Configuration & RA-nWG@30 (Ceiling) \\
\hline
1 & voyage-3.5 (2048d) + rerank-2.5 (K=100) & 0.828 (0.898) \\
2 & voyage-3.5 (512d)  + rerank-2.5 (K=100) & 0.824 (0.892) \\
3 & voyage-3.5 (1024d) + rerank-2.5 (K=100) & 0.819 (0.889) \\
4 & voyage-3.5 (1024d) + rerank-2.5 (K=200) & 0.818 (0.936) \\
5 & voyage-3.5 (512d)  + rerank-2.5 (K=50)  & 0.817 (0.847) \\
\hline
\end{tabular}
\endgroup
\end{table}

\textit{Observation.} For dense-reranked systems, \(K=50\) yields the best (or tied-best) @10 quality; \(K=100\) lifts @30 with little change at @10. 512d/1024d often match 2048d at @10, while 2048d wins some deep-\(K\) cases (see \S5.4 for latency considerations at high \(K\)). Full per-configuration tables appear in Appendix~A.7--A.8. Dense PROC ceiling tables appear in Appendix~A.9 and are the “Ceiling” values used above.

% ---------- 5.x Conclusion (inline under §5.2) ----------
\paragraph*{Conclusion.}
We summarize both the deployment-facing takeaway and the mechanism we infer from the ceilings.
\begin{sloppypar}
Across all dense-reranked runs, \emph{rerank-2.5}\allowbreak{} is the safer choice:
it consistently outperforms \emph{rerank-2.5-lite}\allowbreak{}, and no “lite” variant
reaches the top-5. For \textit{shallow quality (@10)}, the best outcomes come from
\textit{Dense $\rightarrow$ rerank-2.5} with $K_{\text{pool}}{=}50$. Pushing the pool
larger does raise the dense \textit{ceiling} \allowbreak{}(the PROC within that pool),
but it does not lift actual @10---those extra candidates mostly introduce \textit{hard
negatives} that the reranker must separate from near-semantic “cousins.” For
\textit{deep quality (@30)}, $K_{\text{pool}}{=}100$ is the sweet spot: it increases
the ceiling enough to matter while keeping distractors in check, yielding better
RA-nWG/N-Recall than 50 without the precision drag (and latency hit) we observe
at 150--200. By dimension, \textit{512d and 1024d} are effectively tied at @10,
while \textit{2048d} occasionally wins at @30; pick \textit{1024d} as a default and
move to \textit{2048d} only if deep-$K$ matters and latency budgets allow. The
\textit{Hybrid+\allowbreak Rerank} reference beats dense-reranked because its
\textit{pool coverage} is stronger---i.e., that gain is primarily \textit{retrieval
headroom}, not just better ordering. This aligns with prior work showing neural
re-rankers’ gains are contingent on strong candidate pools \citep{Nogueira2019BERTPR,Thakur2021BEIR}.
\end{sloppypar}

Mechanistically, the results separate \textit{ordering headroom} from \textit{retrieval headroom}. At \textit{@10}, dense-reranked configurations realize roughly \textit{83--89\%} of their dense ceilings; at \textit{@30}, utilization rises to around \textit{$\sim$92\%} overall (with a \textit{$\sim$96.5\%} high for $K_{\text{pool}}{=}50$ and a \textit{$\sim$87\%} dip for $K_{\text{pool}}{=}200$). In other words, reranking is relatively more effective at deeper cutoffs, where there are simply \textit{more true positives to elevate}, and RA-nWG's rarity weighting softens the penalty from a few residual distractors. This also explains the operating points: \textit{$K_{\text{pool}}{=}50$} is ideal for @10 because the ``best 10'' are usually already present and adding more candidates mainly injects hard negatives that compress margins; \textit{$K_{\text{pool}}{=}100$} wins at @30 because the \textit{ceiling lifts} enough to expose additional relevant items without drowning the reranker in near-misses. The \textit{lite} reranker underperforms because the token savings come at the cost of \textit{weaker margins on hard negatives}, and those misses show up first at @10 where precision pressure is highest. Two low-effort improvements follow directly: \textit{dedupe/near-duplicate suppression} before rerank to thin hard negatives, and \textit{light lexical boosts (e.g., BM25 features) for rare-signal passages} to align with RA-nWG's rarity weighting. Longer-term, a \textit{dynamic $K_{\text{pool}}$}---e.g., 50 for ``easy'' queries and 100 for ``hard,'' triggered by retrieval uncertainty---preserves @10 while lifting @30 without paying 200-scale costs.

% ---------- §5.3 Quantization and ANN effects ----------
\subsection{Quantization and ANN effects}
HNSW vs.\ exact flat cosine and int8 quantization (baseline: \texttt{voyage-3.5}, 1024d; $n{=}50$ queries):

\begin{table}[H]
\centering
\caption{ANN/quantization comparison at 1024d. Columns @15/@25/@30 and Emb Lat.\ removed; deltas vs.\ flat shown in parentheses.}
\label{tab:ann-quant}
\begingroup\small
\setlength{\tabcolsep}{6pt}
\renewcommand{\arraystretch}{1.15}
\begin{adjustbox}{max width=\linewidth}
\begin{tabular}{l
c c c
c c c
r}
\hline
Setup &
\makecell[c]{N-Recall\\4+\\@10} & \makecell[c]{N-Recall\\5\\@10} & \makecell[c]{RA-\\nWG\\@10} &
\makecell[c]{N-Recall\\4+\\@20} & \makecell[c]{N-Recall\\5\\@20} & \makecell[c]{RA-\\nWG\\@20} &
\makecell[c]{Median\\Lat.\\(ms)} \\
\hline
\makecell[l]{flat-cos-1024}
& 0.592 & 0.596 & 0.566
& 0.636 & 0.677 & 0.625
& 47.6 \\
\makecell[l]{hnsw-f32}
& 0.592 (+0.0\%) & 0.596 (+0.0\%) & 0.566 (+0.0\%)
& 0.594 (+0.8\%) & 0.616 (-0.7\%) & 0.580 (+0.6\%)
& 35.4 (-25.6\%) \\
\makecell[l]{hnsw-int8-fast50}
& 0.521 (-12.0\%) & 0.507 (-14.9\%) & 0.497 (-12.1\%)
& 0.526 (-17.3\%) & 0.555 (-18.1\%) & 0.522 (-16.5\%)
& 35.0 (-26.3\%) \\
\hline
\end{tabular}
\end{adjustbox}
\endgroup
\end{table}

With $n{=}50$ queries and \texttt{voyage-3.5} (1024d), the pattern is unambiguous. Moving from exact \textit{flat cosine} to \textit{HNSW (float32)} \citep{Malkov2020HNSW} yields a $\sim$26\% drop in retrieval latency with no measurable quality loss at @10 and @20 and only tiny, non-systematic wiggles at deeper cutoffs. In contrast, under the same aggressive search settings, \textit{int8 quantization} trades away $\sim$8--18\% of quality for virtually no additional speedup beyond HNSW-F32: retrieval time is essentially the same as float32 HNSW, while RA-nWG and N-Recall$_{4+}/_{5}$ degrade across all $K$. Given final LLM inference dominates the end-to-end budget, those few milliseconds saved at the retriever cannot compensate for the quality loss. Comparable memory-aware approaches such as product/optimized product quantization often preserve recall better than na\"ive int8 when tuned for high-recall ANN \citep{Jegou2011PQ,Ge2013OPQ,Jacob2018INT8}.

Mechanistically, this fits the geometry. \textit{HNSW-F32} preserves full-precision neighborhoods; any approximation error lives in the tail of the candidate set, which set-based metrics largely tolerate---especially at @10--@20, where the best evidence is consistently surfaced. \textit{Int8} changes the vector space itself: margins between near neighbors shrink, raising \textit{hard-negative confusion} (semantically close distractors). Because \textit{RA-nWG} is rarity-aware, replacing even a handful of high-utility passages with near misses is disproportionately costly.

\paragraph*{Memory considerations (order-of-magnitude, implementation-agnostic).}
\begin{itemize}
  \item \textbf{Vector payload.} For dimension $d{=}1024$: \textit{float32}: $4\times d{=}4096$~B per vector $\approx$ 3.8~GiB per million; \textit{int8}: $1\times d{=}1024$~B $\approx$ 0.95~GiB per million. Thus, int8 saves $\sim$2.9~GiB per million (about $4\times$ compression).
  \item \textbf{Graph overhead (HNSW).} Independent of quantization, neighbor links and metadata add $\sim$0.15--0.6~GiB per million nodes (typical $M{\approx}16$--32 and 32/64-bit IDs). This component does not shrink when you quantize vectors to int8.
  \item \textbf{Net effect.} For 10M vectors: vectors $\sim$38~GiB (f32) vs.\ $\sim$9.5~GiB (int8); adding HNSW graph (e.g., +1.5--6~GiB) yields roughly 10--15~GiB (int8) vs.\ 40--44~GiB (f32). That is a substantial RAM reduction, but with the observed 8--18\% quality loss and negligible speed gain, take it only under hard memory constraints.
\end{itemize}

\paragraph*{Practical guidance.}
Default to \textit{HNSW-F32} and tune recall (\texttt{efSearch}) until the delta vs.\ flat at @10/@30 is $\le$1--2\%; further tuning mostly burns latency. Treat \textit{int8} as a memory lever only: if RAM is the bottleneck, quantify the benefit with the back-of-the-envelope above and then re-check \% of PROC (ceiling) to confirm you are not spending more in quality than you saved in hardware. If you do need stronger compression with milder quality loss, consider \textit{learned/product quantization} with recall-aware search rather than blunt int8.

\subsection{Latency scaling and ``efficiency''}

We ran the reranker three times at different times of day and report median reranker latency, averaged across 50 queries $\times$ models $\times$ dimensions $\times$ reranker tier. Under this protocol, latency scales primarily with $K$ (candidate count), not with embedding dimensionality. For \texttt{rerank-2.5-lite}, medians increase smoothly from roughly $340$--$405$ ms at $K{=}50$ to $\sim 640$--$715$ ms at $K{=}200$. For \texttt{rerank-2.5}, most configurations follow a similar trend, rising from $\sim 331$--$339$ ms ($K{=}50$) to $\sim 580$--$670$ ms ($K{=}150$--$200$).

A notable exception is \texttt{voyage-3.5} at high $K$: several 3.5 variants exhibit a latency discontinuity at $K \ge 150$ (most obvious at $K{=}200$), with medians jumping to $\sim 2.7$--$3.0$ s, whereas \texttt{voyage-3-large} remains stable around $\sim 0.6$--$0.7$ s in the same regime. Because $512$d vs.\ $2048$d track closely elsewhere ($\le$~40 ms differences at $K \le 100$), this pattern is unlikely to be driven by vector size. A more plausible explanation is provider-side, implementation-dependent non-linearities (e.g., batching thresholds, context fragmentation, throttling, or cache/path differences) specific to certain \texttt{voyage-3.5} configurations. We therefore regard the $K \ge 150$ spike on \texttt{voyage-3.5} as an anomalous behavior that warrants provider investigation rather than as an inherent property of dimensionality or stack design.

\emph{Implications for practitioners.} To keep median latency under $\sim 0.5$ s, set $K \le 100$ for either reranker tier. As complementary mitigations, reduce per-candidate tokens, deduplicate near-duplicates before rerank, or adopt dynamic $K$ (e.g., $50$ for ``easy'' queries; $100$ for ``hard'') to stay within an SLO.

\emph{On ``efficiency'' as a scalar.} Our $\text{Efficiency}=\text{Avg(quality)}/\text{median reranker latency (s)}$ is a handy screening heuristic. Moreover, single weighted-sum scalars are known to obscure Pareto trade-offs \citep{DasDennis1997WSumDrawbacks}. But it is insensitive to SLOs and systematically favors small $K$, precisely where dense-reranked systems already look similar at @10. More importantly, in typical RAG deployments the end-to-end budget is dominated by final LLM inference, with embedding and reranking contributing a smaller (but $K$-sensitive) share. This is consistent with retrieval-augmented generation pipelines where retrieval cost is a pre-inference stage \citep{Lewis2020RAG}. A single reranker-only ratio therefore overstates the operational relevance of small latency deltas at retrieval time. In the main text, we replace the scalar with SLO-conditioned frontiers (e.g., $\le 350$ ms, $\le 500$ ms), include tail latency (p95), and report \% of PROC to separate ordering gains from pool coverage. Tail behavior is operationally critical \citep{Dean2013TailAtScale}. Where a summary number is useful, we recommend two stage-aware variants: (i) a retrieval-local efficiency (same definition as above, for tuning $K$ and ANN knobs), and (ii) an end-to-end efficiency that divides Avg(quality) by total median latency (embedding $+$ retrieval $+$ rerank $+$ final inference) under a fixed prompt budget. Finally, we add marginal analyses---$\Delta$RA-nWG\;/\;$\Delta$ms (e2e) when increasing $K$---to show where additional candidates cease to pay for themselves.

\emph{Presentation choice.} Because of (i) the non-stationarity introduced by time-of-day runs, (ii) the provider-specific high-$K$ discontinuity on \texttt{voyage-3.5}, and (iii) the limited diagnostic value of a single ``efficiency'' scalar, we move the detailed latency tables (median reranker latency for \texttt{rerank-2.5} and \texttt{rerank-2.5-lite} across model/dimension and $K$) to the Appendix and reference them from this section. The main paper retains only (a) the methodological summary above and (b) SLO-anchored recommendations.

\textit{Appendix tables referenced:} A.7 (\texttt{rerank-2.5}: $K \in \{50,100,150,200\}$) and A.8 (\texttt{rerank-2.5-lite}: $K \in \{50,100,150,200\}$), ``Median Reranker Latency (ms) and Metrics vs.\ $K$, by model and dimension.''
\subsection{Efficiency leaderboard and scenario matrix}

Representative CLQ scenarios (priced at 500 tokens/doc; quality and latency measured). All costs shown are per 1,000 queries and cover the rerank call only (tokens counted as query+doc per candidate):

\begin{table}[H]
\centering
\caption{Representative CLQ scenarios with costs per 1{,}000 queries (rerank call only; 500 tokens/candidate).}
\label{tab:clq-scenarios}
\begingroup\small
\setlength{\tabcolsep}{6pt}
\renewcommand{\arraystretch}{1.15}
\resizebox{\linewidth}{!}{%
\begin{tabular}{l l l r r c c c c}
\hline
Scenario & Model (Dim) & Reranker & K & \makecell[c]{Cost} & \makecell[c]{Latency\\ms} &
\colNR{4+}{10} & \colRA{10} & \makecell[c]{RA-nWG\\@30} \\
\hline
Baseline            & voyage-3.5 (1024d)      & rerank-2.5      &  50 & \$1.25 &  332.9 & 0.835 & 0.804 & 0.810 \\
Cost saver          & voyage-3.5-lite (1024d) & rerank-2.5-lite &  50 & \$0.50 &  403.8 & 0.710 & 0.692 & 0.732 \\
Quality push        & voyage-3.5 (2048d)      & rerank-2.5      & 100 & \$2.50 &  478.1 & 0.815 & 0.791 & 0.828 \\
Efficient small-dim & voyage-3.5 (512d)       & rerank-2.5      & 100 & \$2.50 &  483.1 & 0.822 & 0.793 & 0.824 \\
High-K check        & voyage-3.5 (1024d)      & rerank-2.5      & 200 & \$5.00 & 2931.1 & 0.815 & 0.792 & 0.818 \\
\hline
\end{tabular}}
\endgroup
\end{table}

\emph{Pricing note.} Costs use 500 tokens/candidate for comparability. The measured mean was \(\sim 515\) tokens/candidate (see §3.1), which would raise per-1k-query rerank costs by \(\sim 3\%\).
\section{Discussion and Conclusions}

RAG should be evaluated as \emph{set consumption}, not rank browsing. In practice this means reporting the \emph{trio} of \emph{RA-nWG@}\(K\) (rarity-aware, per-query--normalized utility), \emph{N-Recall}\(_{4+}\)\emph{@}\(K\) (coverage of good evidence), and \emph{Harm@}\(K\) when the generator is brittle or order-sensitive. This combination aligns with how context is actually used---LLMs ingest a set under a fixed prompt budget---so scores remain interpretable at fixed \(K\) and token limits. To decide which knob to turn next, we rely on \emph{pool-restricted oracle ceilings (PROC)} and the realized \(\%\)PROC: when PROC is low, the ceiling itself is the problem and you should improve retrieval (hybridize dense+BM25, tune ANN, add rewriting/denoising); when PROC is high but realized \(\%\)PROC lags, ordering is the bottleneck and you should strengthen reranking and pre-rerank cleanup (near-duplicate suppression, shorter chunks, cleaner metadata).

Operationally, we advocate \emph{dynamic-\(K\) routing}: default \(K=50\) for most queries, and automatically escalate to \(K=100\) only when uncertainty signals trigger---e.g., small dense-cosine margins among top candidates, high reranker entropy, or \(\Delta\)-diagnostic drops (names/noise). This preserves @10 precision, lifts @30 recall, and keeps retrieval p50 within typical SLAs. The most reliable default stack is \emph{Hybrid+Rerank}: build the pool with \emph{Dense+BM25 (RRF-100)} to raise the ceiling, then apply \emph{rerank-2.5} to realize it; reserve \emph{2.5-lite} for hard cost caps or recall-heavy, precision-tolerant settings. Two low-effort levers compound these gains: deduplicate near-duplicates before rerank and trim chunk length/metadata bloat, because with \(\sim 500\) tokens per candidate, \(K\) directly multiplies both rerank and generation spend and slows first-token latency.

Finally, \emph{names and noise} need explicit guardrails. The proper-name signal is real and useful; identity-destroying changes (hard masks, gibberish, near-miss edits) collapse it, whereas case, order, and diacritics are largely benign. Conversational padding can depress cosine by \(20\text{--}40\%\) (typically worse in FR), so make \emph{rewrite/denoise} the default and enforce \emph{Unicode hygiene} to stabilize multilingual retrieval. Methodologically, \emph{RA-nWG} is order-free and redundancy-agnostic by design; where distractors or within-prompt order matter, pair it with a \emph{novelty discount} and optionally \emph{UDCG/Harm@}\(K\), and track \emph{Acc\(|\)Hit} to validate the recall-first premise. To keep these conclusions portable and auditable, ship results with \emph{rag-gs} manifests, configs, and per-run workspaces, including PROC/\(\%\)PROC and SLOs---so CLQ claims can be reproduced across stacks and over time.

\section*{Acknowledgments}
This work was conducted independently, largely during personal time over the holidays, without external funding or institutional compute. I’m grateful to the maintainers of the open-source tooling used throughout (Elasticsearch/BM25, HNSW indices).

\section*{Limitations}
\textbf{Model coverage (retrievers).} Most experiments use Voyage AI embedders, with a small number of OpenAI variants for comparison. I actually have more OpenAI embedding results than shown here, but I didn’t have time to clean and integrate them. Major families (E5/BGE/Instructor, Cohere, Jina, Snowflake Arctic, Nomic, mixed-breadth sentence-transformers) were not included.

\textbf{Corpora \& questions.} The study reflects a production RAG over a science-paper corpus, using a hybrid dense+BM25 setup. Queries are real-world and span a broad spectrum: some are very specific with a single decisive answer; others are broad and require coverage across many relevant passages. To isolate variables, I intentionally excluded graph-style / multi-hop aggregation questions (e.g., “How many papers has author X written about subject Y?”) that would require Graph-RAG (counts, joins, entity resolution).

\textbf{Distractor brittleness across LLMs (next steps).} An open question is distractor sensitivity across model generations (e.g., Mistral-70B vs newer GPT-5 family models). A proper follow-up should quantify this.

\textbf{Anecdotal production check.} In a quick relevance spot-check on the production hybrid system, domain experts rated how well answers matched the question/context: mean 4.5/5 (SD \(\approx 0.5\)) over 50 responses. Notes: ratings were by a small expert panel (\(N=2\), inter-rater \(\kappa = 0.82\)), so these are expert quality judgments—not a measure of end-user satisfaction (no SUS/CSAT/NPS collected).

AI tool was used to assist with translation.

\bibliography{references}

\appendix
\section{Appendix}

\subsection{Best Efficiency (Performance/Latency) Leaderboard (A.1)}

\begin{table}[H]
\centering
\caption{Best Efficiency (Performance/Latency) Leaderboard}
\label{tab:app-a1-efficiency}
\begingroup\small
\setlength{\tabcolsep}{8pt}
\renewcommand{\arraystretch}{1.15}
\begin{tabular}{r l c c c}
\hline
Rank & Configuration                              & Efficiency & Avg Performance & Latency (ms) \\
\hline
1    & voyage-3.5 (1024d) + rerank-2.5 (K=50)     & 2.454      & 0.817           & 332.9        \\
2    & voyage-3.5 (512d) + rerank-2.5 (K=50)      & 2.426      & 0.818           & 337.2        \\
3    & voyage-3.5 (2048d) + rerank-2.5 (K=50)     & 2.397      & 0.812           & 338.8        \\
4    & voyage-3-large (1024d) + rerank-2.5 (K=50) & 2.362      & 0.782           & 330.9        \\
5    & voyage-3.5 (512d) + rerank-2.5-lite (K=50) & 2.353      & 0.799           & 339.5        \\
\hline
\end{tabular}
\endgroup
\end{table}

\subsection{Base Condition Margins (\texorpdfstring{$\Delta_{\text{name}}/\Delta_{\text{topic}}/\Delta_{\text{both}}$}{Delta name/topic/both}) (A.2)}

\begin{table}[H]
\centering
\caption{Base Condition Margins}
\label{tab:app-a2-margins}
\begingroup\small
\setlength{\tabcolsep}{12pt}
\renewcommand{\arraystretch}{1.15}
\begin{tabular}{l l c c c}
\hline
Lang & Model      & $\Delta_{\text{name}}$ & $\Delta_{\text{topic}}$ & $\Delta_{\text{both}}$ \\
\hline
EN   & OpenAI 3L  & 0.175  & 0.305   & 0.486  \\
EN   & Voyage 3.5 & 0.160  & 0.298   & 0.464  \\
FR   & OpenAI 3L  & 0.139  & 0.260   & 0.407  \\
FR   & Voyage 3.5 & 0.164  & 0.277   & 0.447  \\
\hline
\end{tabular}
\endgroup
\end{table}

\subsection{Proper-Name vs Topic Signal Ratio (A.3)}

\begin{table}[H]
\centering
\caption{Proper-Name vs Topic Signal Ratio}
\label{tab:app-a3-ratios}
\begingroup\small
\setlength{\tabcolsep}{18pt}
\renewcommand{\arraystretch}{1.15}
\begin{tabular}{l l c}
\hline
Lang & Model      & Name/Topic Ratio \\
\hline
EN   & OpenAI 3L  & 0.574            \\
EN   & Voyage 3.5 & 0.537            \\
FR   & OpenAI 3L  & 0.535            \\
FR   & Voyage 3.5 & 0.592            \\
\hline
\end{tabular}
\endgroup
\end{table}

\subsection{Conversational-noise cosine drops (EN/FR; large vs.\ small) (A.4)}

\begin{table}[H]
\centering
\caption{Cosine similarity by noise level}
\label{tab:app-a4-cosine}
\begingroup\small
\setlength{\tabcolsep}{14pt}
\renewcommand{\arraystretch}{1.15}
\begin{tabular}{l l r r}
\hline
Language & Noise Level        & Cosine --- Large & Cosine --- Small \\
\hline
French   & 0 --- No noise       & 0.818 & 0.936 \\
         & 2 --- Moderate noise  & 0.653 & 0.757 \\
         & 4 --- High noise      & 0.522 & 0.559 \\
English  & 0 --- No noise       & 0.828 & 0.908 \\
         & 2 --- Moderate noise  & 0.749 & 0.806 \\
         & 4 --- High noise      & 0.593 & 0.634 \\
\hline
\end{tabular}
\endgroup
\end{table}

\begin{table}[H]
\centering
\caption{Average drop from clean to high-noise (absolute cosine difference)}
\label{tab:app-a4-drops}
\begingroup\small
\setlength{\tabcolsep}{22pt}
\renewcommand{\arraystretch}{1.15}
\begin{tabular}{l r r}
\hline
Language & $\Delta$ --- Large & $\Delta$ --- Small \\
\hline
French   & $-0.296$ & $-0.377$ \\
English  & $-0.235$ & $-0.274$ \\
\hline
\end{tabular}
\endgroup
\end{table}

\noindent
These drops correspond to $\sim 28\text{--}40\%$ relative to the clean condition (FR-Large: $36\%$, FR-Small: $40\%$, EN-Large: $28\%$, EN-Small: $30\%$), consistent with the $20\text{--}40\%$ headline in \S4.3.

\subsection{Measured Mean Tokens Per Candidate (A.5)}

\begin{table}[H]
\centering
\caption{Measured mean tokens per candidate (query + chunk + metadata).}
\label{tab:a5-mean-tokens}
\begingroup\small
\setlength{\tabcolsep}{10pt}
\renewcommand{\arraystretch}{1.15}
\begin{tabular}{l r r r r}
\hline
Reranker Model  & K    & Mean Tokens & Std Dev & Configs \\
\hline
rerank-2.5      & 50   & 516.0       & 2.3     & 5       \\
rerank-2.5      & 100  & 514.1       & 1.9     & 5       \\
rerank-2.5      & 150  & 513.0       & 2.1     & 5       \\
rerank-2.5      & 200  & 512.9       & 1.5     & 5       \\
rerank-2.5-lite & 50   & 516.0       & 2.3     & 5       \\
rerank-2.5-lite & 100  & 514.1       & 1.9     & 5       \\
rerank-2.5-lite & 150  & 513.0       & 2.1     & 5       \\
rerank-2.5-lite & 200  & 512.9       & 1.5     & 5       \\
\hline
\end{tabular}
\endgroup
\end{table}

\subsection{Reranker Latency Summaries (Overview) (A.6)}

\begin{table}[H]
\centering
\caption{Median rerank-2.5 latency (ms) vs.\ K, by model and dimension.}
\label{tab:a6-r25}
\begingroup\small
\setlength{\tabcolsep}{12pt}
\renewcommand{\arraystretch}{1.15}
\begin{tabular}{l r r r r}
\hline
Model                   & K=50  & K=100 & K=150              & K=200              \\
\hline
voyage-3-large (1024d)  & 330.9 & 469.9 & 582.2              & 667.9              \\
voyage-3.5 (1024d)      & 332.9 & 494.1 & 2720.7$^{\dagger}$ & 2931.1$^{\dagger}$ \\
voyage-3.5 (2048d)      & 338.8 & 478.1 & 751.1              & 2970.7$^{\dagger}$ \\
voyage-3.5 (512d)       & 337.2 & 483.1 & 925.7              & 2904.2$^{\dagger}$ \\
voyage-3.5-lite (1024d) & 330.8 & 476.0 & 615.6              & 2940.1$^{\dagger}$ \\
\hline
\end{tabular}
\endgroup

\medskip
\raggedright\footnotesize
$^{\dagger}$\,High-K latency discontinuity (suspected provider-side anomaly); see \S5.4 for discussion.
\end{table}

\begin{table}[H]
\centering
\caption{Median rerank-2.5-lite latency (ms) vs.\ K, by model and dimension.}
\label{tab:a6-r25lite}
\begingroup\small
\setlength{\tabcolsep}{12pt}
\renewcommand{\arraystretch}{1.15}
\begin{tabular}{l r r r r}
\hline
Model                   & K=50  & K=100 & K=150 & K=200 \\
\hline
voyage-3-large (1024d)  & 369.3 & 415.5 & 537.3 & 639.2 \\
voyage-3.5 (1024d)      & 352.2 & 413.9 & 515.9 & 644.8 \\
voyage-3.5 (2048d)      & 405.4 & 474.6 & 611.7 & 715.3 \\
voyage-3.5 (512d)       & 339.5 & 418.2 & 557.3 & 694.0 \\
voyage-3.5-lite (1024d) & 403.8 & 411.3 & 530.5 & 695.8 \\
\hline
\end{tabular}
\endgroup
\end{table}

\subsection{Appendix A.8 — Reranker metrics and latency by K (rerank-2.5-lite)}

\noindent\textit{Methodology.} Same protocol as A.7; these appendix tables use the 15-query subsample for latency probing, whereas the main text uses $n=50$ (see \S5.1).

\FloatBarrier

\begin{table}[H]
\centering
\caption{rerank-2.5-lite (Reranker K=50)}
\label{tab:a8-r25lite-k50}
\begingroup
\small
\setlength{\tabcolsep}{5pt}
\renewcommand{\arraystretch}{1.15}
\resizebox{\textwidth}{!}{
\begin{tabular}{l *{9}{r} r}
\hline
Model & N-Recall4+@10 & N-Recall5@10 & RA-nWG@10 & N-Recall4+@20 & N-Recall5@20 & RA-nWG@20 & N-Recall4+@30 & N-Recall5@30 & RA-nWG@30 & Median Reranker Latency (ms) \\
\hline
voyage-3-large (1024d)  & 0.779 & 0.762 & 0.746 & 0.780 & 0.796 & 0.762 & 0.798 & 0.802 & 0.781 & 369.3 \\
voyage-3.5 (1024d)      & 0.799 & 0.772 & 0.767 & 0.791 & 0.821 & 0.784 & 0.814 & 0.824 & 0.799 & 352.2 \\
voyage-3.5 (2048d)      & 0.792 & 0.766 & 0.759 & 0.780 & 0.812 & 0.776 & 0.813 & 0.832 & 0.804 & 405.4 \\
voyage-3.5 (512d)       & 0.799 & 0.779 & 0.769 & 0.793 & 0.837 & 0.792 & 0.817 & 0.848 & 0.811 & 339.5 \\
voyage-3.5-lite (1024d) & 0.710 & 0.732 & 0.692 & 0.711 & 0.780 & 0.713 & 0.737 & 0.776 & 0.732 & 403.8 \\
\hline
\end{tabular}}
\endgroup
\end{table}

\begin{table}[H]
\centering
\caption{rerank-2.5-lite (Reranker K=100)}
\label{tab:a8-r25lite-k100}
\begingroup
\small
\setlength{\tabcolsep}{5pt}
\renewcommand{\arraystretch}{1.15}
\resizebox{\textwidth}{!}{
\begin{tabular}{l *{9}{r} r}
\hline
Model & N-Recall4+@10 & N-Recall5@10 & RA-nWG@10 & N-Recall4+@20 & N-Recall5@20 & RA-nWG@20 & N-Recall4+@30 & N-Recall5@30 & RA-nWG@30 & Median Reranker Latency (ms) \\
\hline
voyage-3-large (1024d)  & 0.772 & 0.766 & 0.749 & 0.751 & 0.786 & 0.744 & 0.801 & 0.821 & 0.792 & 415.5 \\
voyage-3.5 (1024d)      & 0.792 & 0.766 & 0.759 & 0.760 & 0.787 & 0.750 & 0.801 & 0.800 & 0.782 & 413.9 \\
voyage-3.5 (2048d)      & 0.785 & 0.782 & 0.765 & 0.758 & 0.803 & 0.756 & 0.802 & 0.823 & 0.795 & 474.6 \\
voyage-3.5 (512d)       & 0.792 & 0.772 & 0.760 & 0.766 & 0.802 & 0.760 & 0.807 & 0.819 & 0.794 & 418.2 \\
voyage-3.5-lite (1024d) & 0.730 & 0.756 & 0.714 & 0.708 & 0.771 & 0.709 & 0.738 & 0.788 & 0.737 & 411.3 \\
\hline
\end{tabular}}
\endgroup
\end{table}

\begin{table}[H]
\centering
\caption{rerank-2.5-lite (Reranker K=150)}
\label{tab:a8-r25lite-k150}
\begingroup
\small
\setlength{\tabcolsep}{5pt}
\renewcommand{\arraystretch}{1.15}
\resizebox{\textwidth}{!}{
\begin{tabular}{l *{9}{r} r}
\hline
Model & N-Recall4+@10 & N-Recall5@10 & RA-nWG@10 & N-Recall4+@20 & N-Recall5@20 & RA-nWG@20 & N-Recall4+@30 & N-Recall5@30 & RA-nWG@30 & Median Reranker Latency (ms) \\
\hline
voyage-3-large (1024d)  & 0.792 & 0.789 & 0.771 & 0.758 & 0.801 & 0.755 & 0.805 & 0.830 & 0.797 & 537.3 \\
voyage-3.5 (1024d)      & 0.785 & 0.782 & 0.764 & 0.754 & 0.799 & 0.753 & 0.797 & 0.816 & 0.788 & 515.9 \\
voyage-3.5 (2048d)      & 0.779 & 0.776 & 0.759 & 0.751 & 0.793 & 0.749 & 0.794 & 0.810 & 0.784 & 611.7 \\
voyage-3.5 (512d)       & 0.792 & 0.789 & 0.770 & 0.758 & 0.805 & 0.757 & 0.801 & 0.821 & 0.791 & 557.3 \\
voyage-3.5-lite (1024d) & 0.765 & 0.769 & 0.747 & 0.738 & 0.785 & 0.738 & 0.787 & 0.800 & 0.779 & 530.5 \\
\hline
\end{tabular}}
\endgroup
\end{table}

\begin{table}[H]
\centering
\caption{rerank-2.5-lite (Reranker K=200)}
\label{tab:a8-r25lite-k200}
\begingroup
\small
\setlength{\tabcolsep}{5pt}
\renewcommand{\arraystretch}{1.15}
\resizebox{\textwidth}{!}{
\begin{tabular}{l *{9}{r} r}
\hline
Model & N-Recall4+@10 & N-Recall5@10 & RA-nWG@10 & N-Recall4+@20 & N-Recall5@20 & RA-nWG@20 & N-Recall4+@30 & N-Recall5@30 & RA-nWG@30 & Median Reranker Latency (ms) \\
\hline
voyage-3-large (1024d)  & 0.792 & 0.789 & 0.771 & 0.758 & 0.801 & 0.755 & 0.796 & 0.817 & 0.786 & 639.2 \\
voyage-3.5 (1024d)      & 0.785 & 0.782 & 0.764 & 0.754 & 0.799 & 0.753 & 0.803 & 0.813 & 0.792 & 644.8 \\
voyage-3.5 (2048d)      & 0.785 & 0.782 & 0.764 & 0.754 & 0.799 & 0.753 & 0.795 & 0.813 & 0.785 & 715.3 \\
voyage-3.5 (512d)       & 0.792 & 0.789 & 0.770 & 0.758 & 0.805 & 0.757 & 0.799 & 0.819 & 0.788 & 694.0 \\
voyage-3.5-lite (1024d) & 0.772 & 0.769 & 0.751 & 0.744 & 0.785 & 0.741 & 0.794 & 0.800 & 0.782 & 695.8 \\
\hline
\end{tabular}}
\endgroup
\end{table}

\subsection{Appendix A.9 — Ceiling metrics (PROC) by model and candidate pool K (dense pools)}

\noindent\textbf{Definition.} “Ceiling” here means \emph{PROC—Dense-$K_{\text{pool}}$}: the oracle score obtained by perfectly reordering the Dense Top-$K_{\text{pool}}$ for that model. These ceilings are independent of the reranker tier (2.5 vs.\ 2.5-lite). We label sections as “Reranker K=\dots” only to align with the candidate $K$ budgets used in \S5.

\noindent\textbf{How to read.} @10/@20/@30 are evaluation cutoffs within the fixed Dense Top-$K_{\text{pool}}$. Use these tables to compute utilization in \S5.2: \%PROC $=$ Actual / Ceiling. Do not average ceilings across $K$; compare like-for-like $K$ only.

\noindent\textbf{Scope.} These ceilings are for Dense pools. They are not directly comparable to the Hybrid PROC in \S5.1 (“Hybrid RRF-100 $\rightarrow$ Rerank-2.5 $\rightarrow$ Top-50”).

\medskip
\noindent\textit{Ceiling Metrics by Model and Reranker K}

\FloatBarrier

% ---------- K = 50 ----------
\begin{table}[H]
\centering
\caption{Reranker K=50 \; (Ceiling convention: PROC—Dense-$K_{\text{pool}}=50$)}
\label{tab:a9-proc-k50}
\begingroup
\small
\setlength{\tabcolsep}{5pt}
\renewcommand{\arraystretch}{1.15}
\resizebox{\textwidth}{!}{
\begin{tabular}{l *{9}{r}}
\hline
Model & N-Recall4+@10 & N-Recall5@10 & RA-nWG@10 & N-Recall4+@20 & N-Recall5@20 & RA-nWG@20 & N-Recall4+@30 & N-Recall5@30 & RA-nWG@30 \\
\hline
voyage-3-large (1024d)  & 0.919 & 0.860 & 0.875 & 0.839 & 0.846 & 0.829 & 0.839 & 0.826 & 0.813 \\
voyage-3.5 (1024d)      & 0.944 & 0.883 & 0.906 & 0.852 & 0.871 & 0.851 & 0.852 & 0.857 & 0.837 \\
voyage-3.5 (2048d)      & 0.944 & 0.890 & 0.911 & 0.852 & 0.878 & 0.856 & 0.852 & 0.861 & 0.840 \\
voyage-3.5 (512d)       & 0.944 & 0.910 & 0.921 & 0.851 & 0.897 & 0.861 & 0.851 & 0.884 & 0.847 \\
voyage-3.5-lite (1024d) & 0.870 & 0.867 & 0.851 & 0.795 & 0.830 & 0.796 & 0.759 & 0.797 & 0.755 \\
\hline
\end{tabular}}
\endgroup
\end{table}

% ---------- K = 100 ----------
\begin{table}[H]
\centering
\caption{Reranker K=100 \; (Ceiling convention: PROC—Dense-$K_{\text{pool}}=100$)}
\label{tab:a9-proc-k100}
\begingroup
\small
\setlength{\tabcolsep}{5pt}
\renewcommand{\arraystretch}{1.15}
\resizebox{\textwidth}{!}{
\begin{tabular}{l *{9}{r}}
\hline
Model & N-Recall4+@10 & N-Recall5@10 & RA-nWG@10 & N-Recall4+@20 & N-Recall5@20 & RA-nWG@20 & N-Recall4+@30 & N-Recall5@30 & RA-nWG@30 \\
\hline
voyage-3-large (1024d)  & 0.946 & 0.897 & 0.911 & 0.868 & 0.888 & 0.871 & 0.868 & 0.879 & 0.859 \\
voyage-3.5 (1024d)      & 0.982 & 0.917 & 0.942 & 0.892 & 0.909 & 0.899 & 0.892 & 0.904 & 0.889 \\
voyage-3.5 (2048d)      & 0.982 & 0.933 & 0.953 & 0.896 & 0.926 & 0.909 & 0.896 & 0.919 & 0.898 \\
voyage-3.5 (512d)       & 0.976 & 0.923 & 0.942 & 0.895 & 0.914 & 0.903 & 0.895 & 0.910 & 0.892 \\
voyage-3.5-lite (1024d) & 0.893 & 0.900 & 0.887 & 0.837 & 0.897 & 0.852 & 0.837 & 0.879 & 0.836 \\
\hline
\end{tabular}}
\endgroup
\end{table}

% ---------- K = 150 ----------
\begin{table}[H]
\centering
\caption{Reranker K=150 \; (Ceiling convention: PROC—Dense-$K_{\text{pool}}=150$)}
\label{tab:a9-proc-k150}
\begingroup
\small
\setlength{\tabcolsep}{5pt}
\renewcommand{\arraystretch}{1.15}
\resizebox{\textwidth}{!}{
\begin{tabular}{l *{9}{r}}
\hline
Model & N-Recall4+@10 & N-Recall5@10 & RA-nWG@10 & N-Recall4+@20 & N-Recall5@20 & RA-nWG@20 & N-Recall4+@30 & N-Recall5@30 & RA-nWG@30 \\
\hline
voyage-3-large (1024d)  & 0.979 & 0.960 & 0.959 & 0.901 & 0.950 & 0.921 & 0.901 & 0.943 & 0.910 \\
voyage-3.5 (1024d)      & 0.989 & 0.940 & 0.956 & 0.911 & 0.931 & 0.920 & 0.911 & 0.929 & 0.911 \\
voyage-3.5 (2048d)      & 0.989 & 0.947 & 0.964 & 0.907 & 0.939 & 0.923 & 0.907 & 0.937 & 0.914 \\
voyage-3.5 (512d)       & 0.989 & 0.947 & 0.957 & 0.918 & 0.937 & 0.927 & 0.918 & 0.934 & 0.918 \\
voyage-3.5-lite (1024d) & 0.967 & 0.947 & 0.962 & 0.896 & 0.940 & 0.918 & 0.896 & 0.931 & 0.906 \\
\hline
\end{tabular}}
\endgroup
\end{table}

% ---------- K = 200 ----------
\begin{table}[H]
\centering
\caption{Reranker K=200 \; (Ceiling convention: PROC—Dense-$K_{\text{pool}}=200$)}
\label{tab:a9-proc-k200}
\begingroup
\small
\setlength{\tabcolsep}{5pt}
\renewcommand{\arraystretch}{1.15}
\resizebox{\textwidth}{!}{
\begin{tabular}{l *{9}{r}}
\hline
Model & N-Recall4+@10 & N-Recall5@10 & RA-nWG@10 & N-Recall4+@20 & N-Recall5@20 & RA-nWG@20 & N-Recall4+@30 & N-Recall5@30 & RA-nWG@30 \\
\hline
voyage-3-large (1024d)  & 0.979 & 0.960 & 0.959 & 0.905 & 0.950 & 0.923 & 0.905 & 0.946 & 0.913 \\
voyage-3.5 (1024d)      & 1.000 & 0.953 & 0.976 & 0.933 & 0.944 & 0.944 & 0.933 & 0.944 & 0.936 \\
voyage-3.5 (2048d)      & 0.989 & 0.953 & 0.965 & 0.922 & 0.944 & 0.933 & 0.922 & 0.944 & 0.926 \\
voyage-3.5 (512d)       & 0.989 & 0.967 & 0.967 & 0.929 & 0.956 & 0.941 & 0.929 & 0.956 & 0.933 \\
voyage-3.5-lite (1024d) & 0.980 & 0.960 & 0.975 & 0.913 & 0.953 & 0.935 & 0.913 & 0.951 & 0.927 \\
\hline
\end{tabular}}
\endgroup
\end{table}

\FloatBarrier

\end{document}